\documentclass[conference]{IEEEtran}
\IEEEoverridecommandlockouts

\usepackage{cite}
\usepackage{amsmath,amssymb,amsfonts}
\usepackage{amsthm}
\usepackage{graphicx}
\usepackage{subcaption}
\usepackage{textcomp}
\usepackage{xcolor}
\usepackage{xurl}
\usepackage[hidelinks]{hyperref}
\usepackage{algorithm}
\usepackage{algpseudocode}

\definecolor{commentcolor}{RGB}{200,0,180}

\algrenewcommand{\algorithmiccomment}[1]{%
  \hfill{\footnotesize\textcolor{commentcolor}{\textit{// #1}}}%
}

\algnewcommand{\LineComment}[1]{%
  \State {\footnotesize\textcolor{commentcolor}{\textit{// #1}}}%
}

\def\BibTeX{{\rm B\kern-.05em{\sc i\kern-.025em b}\kern-.08em
    T\kern-.1667em\lower.7ex\hbox{E}\kern-.125emX}}

\usepackage{xcolor}
\definecolor{casepink}{RGB}{218,70,145}
\newcommand{\PinkCase}[1]{\Statex \textcolor{brown}{\bfseries #1}}

\begin{document}
\bstctlcite{IEEEexample:BSTcontrol}
\title{BIER-DC: Stateless Multicast for In-Orbit Space Data Centers in LEO Constellations\\
}

\author{
\IEEEauthorblockN{
Mostafa Abdollahi,
Wenjun Yang,
Kousar Malekinasab,
and Jianping Pan
}

\IEEEauthorblockA{
University of Victoria, BC, Canada\\
Email: \{abdolahi68, wenjunyang, kovvy, pan\}@uvic.ca
}

}

\maketitle

\begin{abstract}
Emerging data-intensive space applications generate large volumes of data that must be processed, stored, replicated, and delivered efficiently. However, relying on terrestrial cloud regions can expose satellite data to high delivery latency and loss of timeliness due to intermittent satellite-to-ground visibility and limited access to ground infrastructure. Space Data Centers (SDCs), which organize collaborative satellites into logical compute and storage clusters, offer a promising approach for in-orbit data management and service continuity. In this paper, we first evaluate how satellite clustering with different sizes affects satellite visibility and gateway outage. We then propose BIER-DC, a stateless one-to-many communication protocol tailored for satellite-based SDCs. Unlike terrestrial multicast protocols designed for stable data center fabrics, BIER-DC decouples inter-SDC forwarding from intra-SDC delivery and encodes compact cluster-level forwarding information in packet headers. Comparative evaluations against BIER, BIER-TE, and Yeti show that BIER-DC significantly reduces packet-header length and byte-hop overhead, improving effective goodput and scalability for next-generation LEO space data centers.
\end{abstract}


\section{Introduction}

\par Modern data-intensive Earth-observation (EO) missions, including high-resolution optical, hyperspectral, and SAR sensing, can generate terabyte-scale daily data depending on the payload and observation mode. Recent estimates suggest that global EO collections are growing by roughly 100 PB per year and may exceed 1400 PB by 2030 \cite{wilkinson2024environmental}. Yet data backhaul is limited by satellite-to-ground visibility, ground infrastructure, and variable downlink capacity. After downlink, data may still traverse long terrestrial paths from gateways to remote cloud or data center regions, adding latency and routing inefficiency. This bottleneck is acute in traditional EO architectures, where ground segments may capture as little as 2\% of available hyperspectral observations \cite{denby2023kodan}. Although broadband mega-constellations show that dense gateway deployments improve ground access, \emph{bent-pipe} and terrestrial-cloud models still constrain timely processing and dissemination \cite{denby2020orbital}. Onboard filtering and in-space computing can reduce this dependency. Prior work shows that, for a 1,000-satellite constellation, downlinking only useful data reduces required ground-station sites from 112 to 6, about a 95\% reduction \cite{denby2020orbital}. 

\par To address the limits of terrestrial-cloud and bent-pipe architectures, companies and startups are exploring space-based data centers with high-capability satellites and orbital platforms. Starcloud has launched a satellite with data-center-class GPU compute, Google's Project Suncatcher studies TPU-equipped solar-powered satellites with optical links for in-space AI, and Lonestar Data Holdings has demonstrated space-based storage for disaster recovery and long-term resilience \cite{starcloud2025,nvidia_starcloud2025,google_suncatcher2025,lonestar2025}. These efforts move computation and storage closer to space-generated data. However, space data centers differ from terrestrial ones, which rely on fixed racks, servers, top-of-rack and aggregation switches, and spine or leaf-spine fabrics \cite{greenberg2009vl2,mysore2009portland}. In space, each satellite acts as a rack-like unit with processing, memory, storage, VM or container hosting, and packet forwarding, but without a stable switch-centric fabric. As a result, existing data center communication protocols become inefficient. Terrestrial replication, cache updates, service synchronization, and multicast often rely on switch-resident forwarding state across hierarchical switches \cite{shahbaz2019elmo,diab2022orca}. In space, there are no dedicated top-of-rack or spine switches, and connectivity depends on resource-constrained satellites and dynamic inter-satellite links. Maintaining per-flow, per-group, or per-replica state across many satellites can therefore consume valuable onboard memory, processing, and control bandwidth.

\begin{figure*}[!t]
    \centering

    \begin{minipage}[t]{0.28\textwidth}
        \vspace{0pt}
        \centering
        \includegraphics[width=\linewidth]{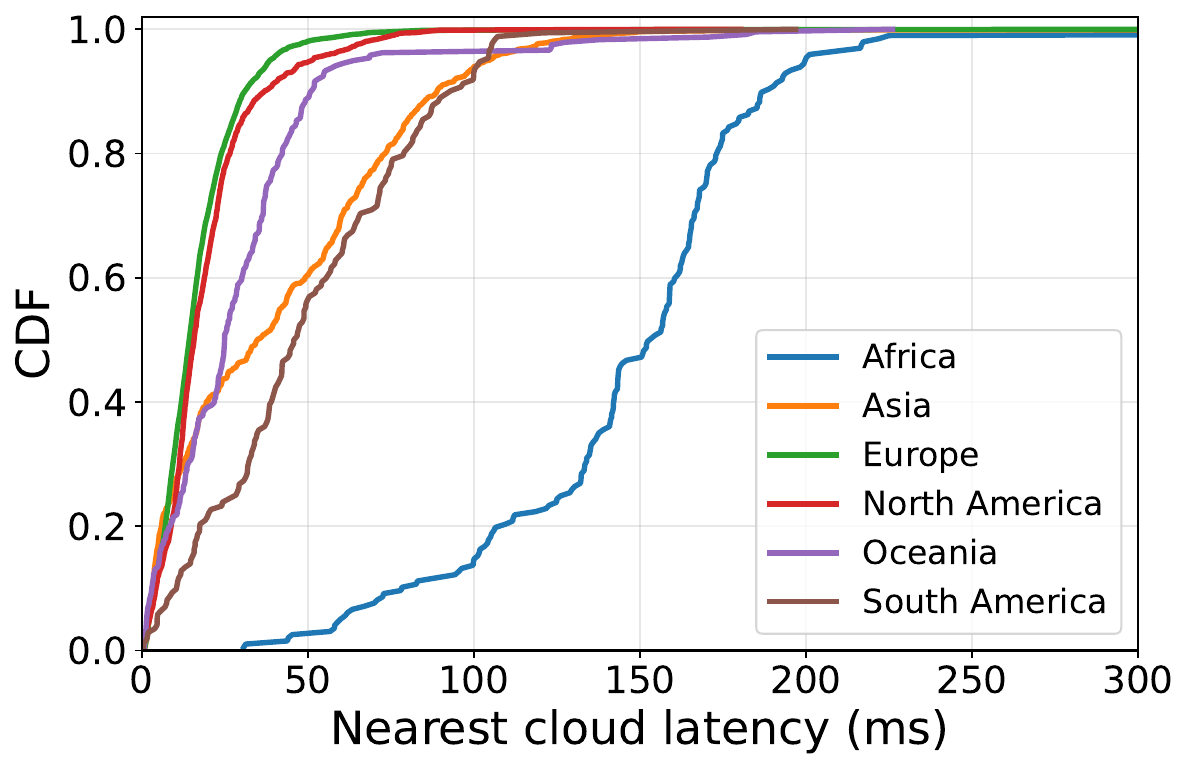}
        \vspace{0mm} 
        \caption{CDF of nearest-cloud access latency by source region.}
        \label{fig:nearest-cloud-latency}
    \end{minipage}
    \hfill
    \begin{minipage}[t]{0.68\textwidth}
        \vspace{0pt}
        \centering
        \captionsetup[subfigure]{labelformat=parens,labelsep=none}

        \begin{subfigure}[t]{0.48\linewidth}
            \centering
            \includegraphics[width=\linewidth]{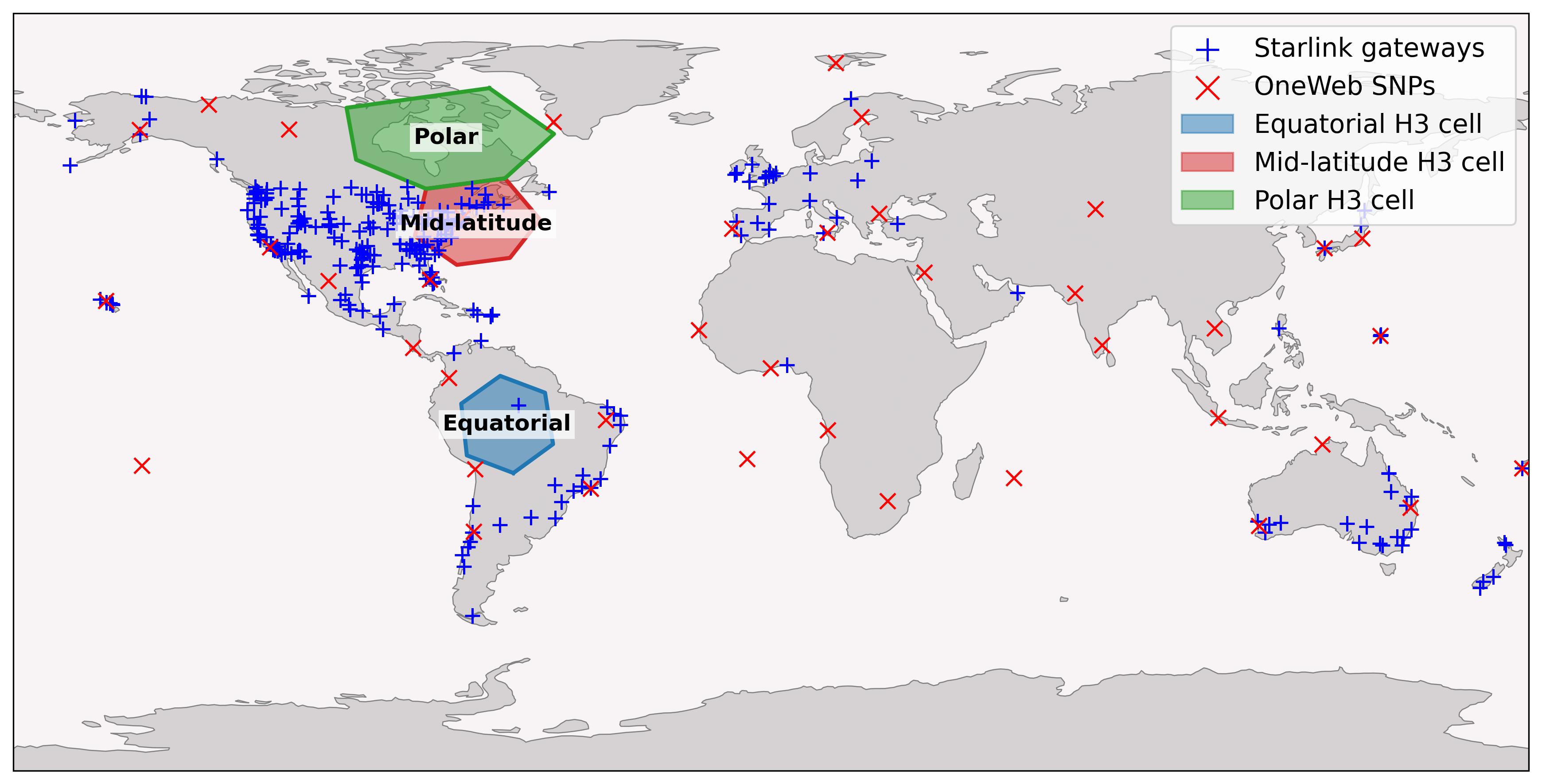}
            \caption{}
            \label{fig:h3-gateway-map}
        \end{subfigure}
        \hfill
        \begin{subfigure}[t]{0.50\linewidth}
            \centering
            \includegraphics[width=\linewidth]{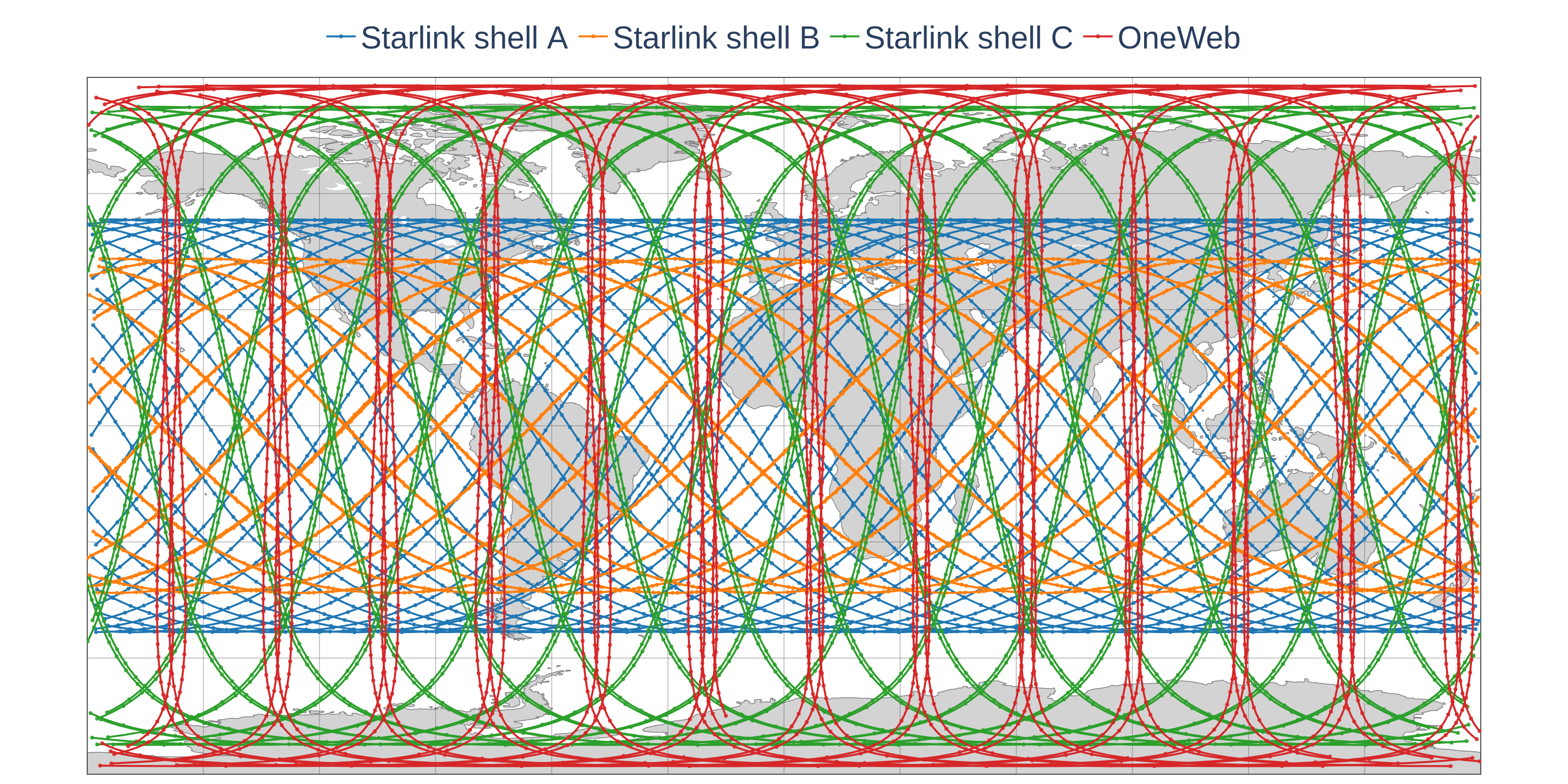}
            \caption{}
            \label{fig:orbital-tracks}
        \end{subfigure}

        \caption{TLE-based analysis setup: (a) selected H3 service cells and ground sites, and (b) representative ground tracks of OneWeb and Starlink shells A--C.} 
        \label{fig:orbital-setup}
    \end{minipage}

    \vspace{-5mm}
\end{figure*}

\par Space data centers require one-to-many communication for replication, fleet-wide management, and distributed learning. As in terrestrial systems that replicate data across data centers and racks for availability \cite{corbett2013spanner}, replication is critical in space to tolerate radiation effects, solar events, and hardware anomalies. Multicast supports fleet-wide firmware, software, configuration, and security updates \cite{sunter2016firmware}, as well as satellite federated learning, where intra-orbit ISLs with partial in-network aggregation achieve a 29$\times$ training speed-up and an 8$\times$ reduction in parameter-server traffic \cite{razmi2022board}. However, existing protocols scale poorly in orbit. IP multicast requires per-flow state and frequent control updates \cite{rfc7761,rfc3569}, while BIER, BIER-TE, and Yeti incur headers that grow with receivers, forwarding adjacencies, or network structure \cite{rfc8279,rfc9262,diab2022yeti}. Terrestrial schemes such as Elmo assume stable Clos topologies and programmable switches \cite{shahbaz2019elmo,singh2015jupiter,xia2016survey}, which do not match dynamic ISL networks. Thus, space data centers need a lightweight stateless multicast design whose header overhead is decoupled from constellation scale. 

\par To address the limitations of existing multicast schemes in orbital environments, we carry out our study in three steps. First, we show that single satellites are insufficient as standalone SDC units. Similar to racks in terrestrial data centers, satellite clusters provide a logical unit for organizing in-orbit compute, storage, and services. This avoids inefficient VM/service communication over individual satellites with changing visibility, gateway access, and ISL connectivity. Since larger clusters improve service continuity and gateway reachability but increase intra-cluster delay, coordination cost, and synchronization overhead, we model SDC formation as a \emph{topology-aware cluster-size selection problem}. Accordingly, the clustering analysis focuses on identifying a practical SDC size and organization for cooperative satellite computing and communication. Intra-cluster latency is used as the primary clustering metric, reflecting the frequent local coordination and data exchange among satellites within an SDC.

\par Second, we design \textsc{BIER-DC} for stateless inter-SDC multicast. Each SDC is assigned a compact cluster identifier, so multicast routing is performed between SDCs rather than individual satellites, links, or VM endpoints. This matches SDC workloads that require frequent one-to-many cluster communication for replication, service updates, AI parameter dissemination, and VM/service coordination. To avoid per-flow multicast state and large satellite/VM-level headers in dynamic LEO networks, \textsc{BIER-DC} encodes destination SDC IDs and selected ingress border satellites in the packet header, while intra-SDC delivery is handled locally through border satellites and topology knowledge.

\par Third, we evaluate \textsc{BIER-DC} over representative LEO constellations using simulation and NetSatBench-based emulation, where each satellite runs as a Docker container and multicast data is transferred using UDP packets. This allows us to measure header overhead, byte-hop overhead, scalability, and the impact of header growth on payload capacity, packet delivery, and goodput. Compared with existing multicast schemes, SDC-level encoding keeps headers compact, preserves more payload space, improves goodput, and maintains \emph{stateless one-to-many forwarding}. The remainder of this paper is organized as follows.
Sec.~\ref{sec:problem} establishes the motivation and context for this work.
Based on these observations, Sec.~\ref{sec:leo-clustering} formulates the topology-aware selection of SDC cluster sizes, which informs the design of \textsc{BIER-DC} presented in Sec.~\ref{sec:bier-dc-design}.
Sec.~\ref{sec:evaluation} then evaluates the proposed design.
Finally, Sec.~\ref{related_works_section} reviews related work, and Sec.~\ref{sec:conclusion} concludes the paper.
\begin{figure*}[t]
    \centering

    \begin{subfigure}[t]{0.32\textwidth}
        \centering
        \includegraphics[width=\linewidth]{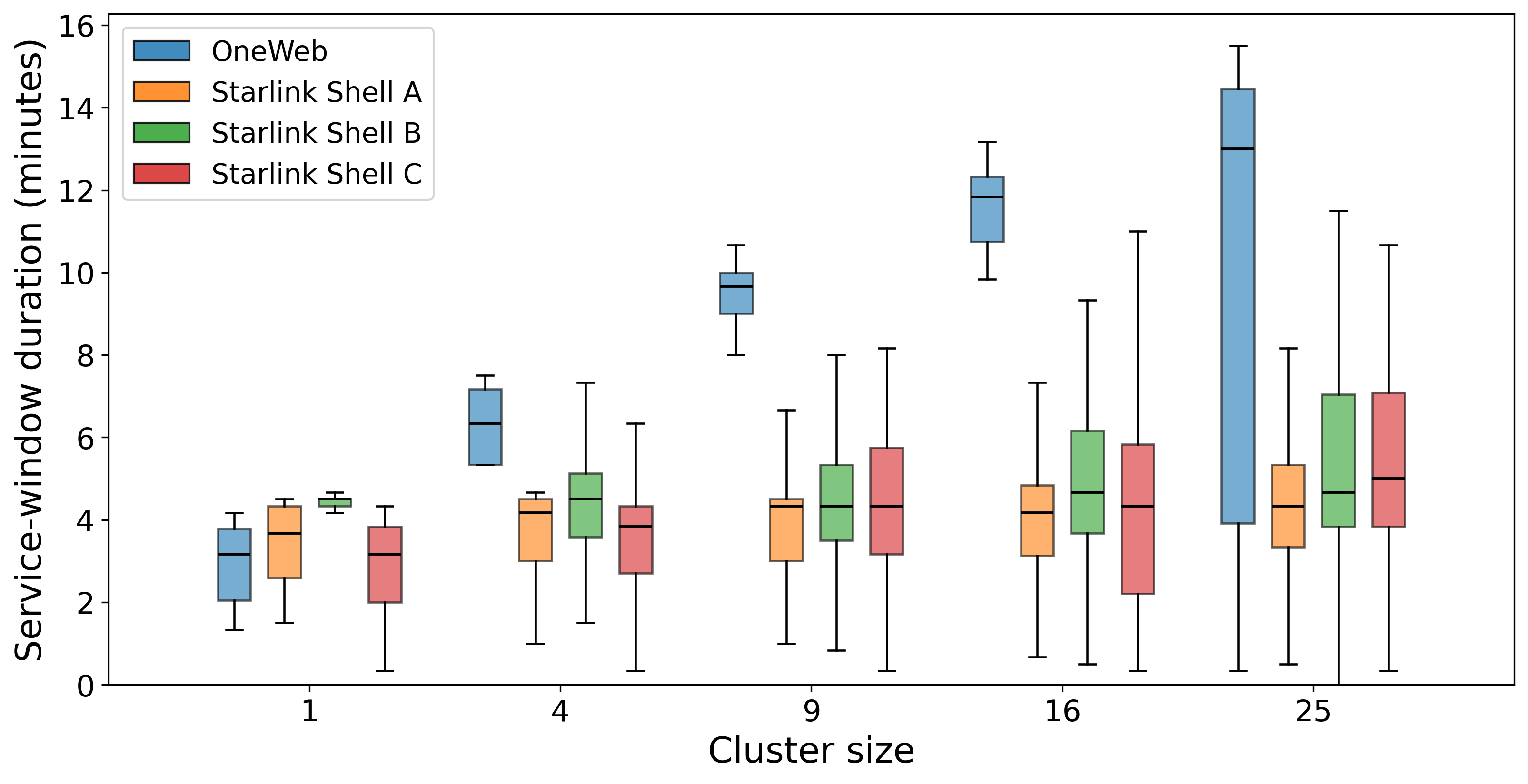}
        \caption{Equatorial.}
        \label{fig:service-window-equatorial}
    \end{subfigure}
    \hfill
    \begin{subfigure}[t]{0.32\textwidth}
        \centering
        \includegraphics[width=\linewidth]{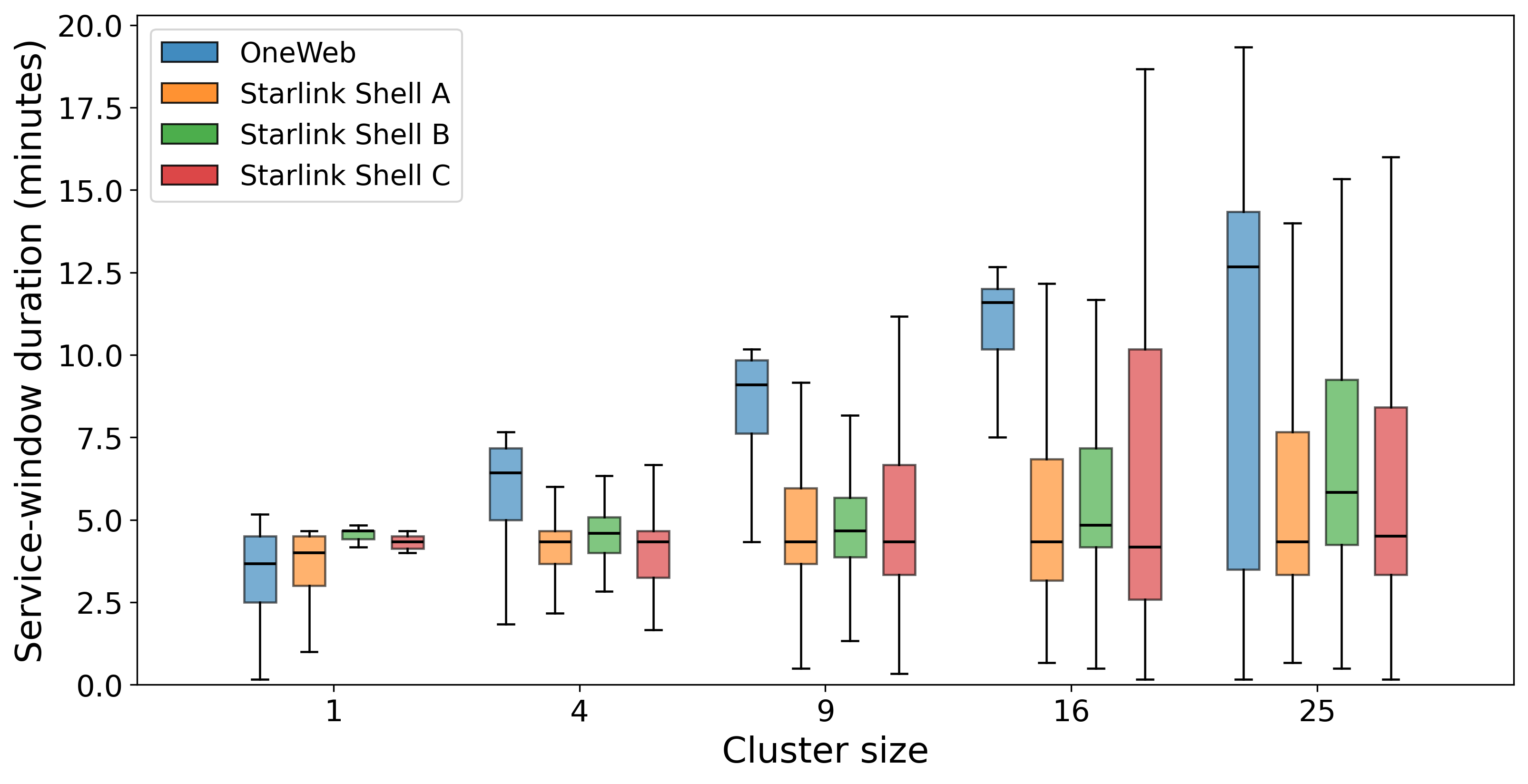}
        \caption{Mid-latitude.}
        \label{fig:service-window-midlatitude}
    \end{subfigure}
    \hfill
    \begin{subfigure}[t]{0.32\textwidth}
        \centering
        \includegraphics[width=\linewidth]{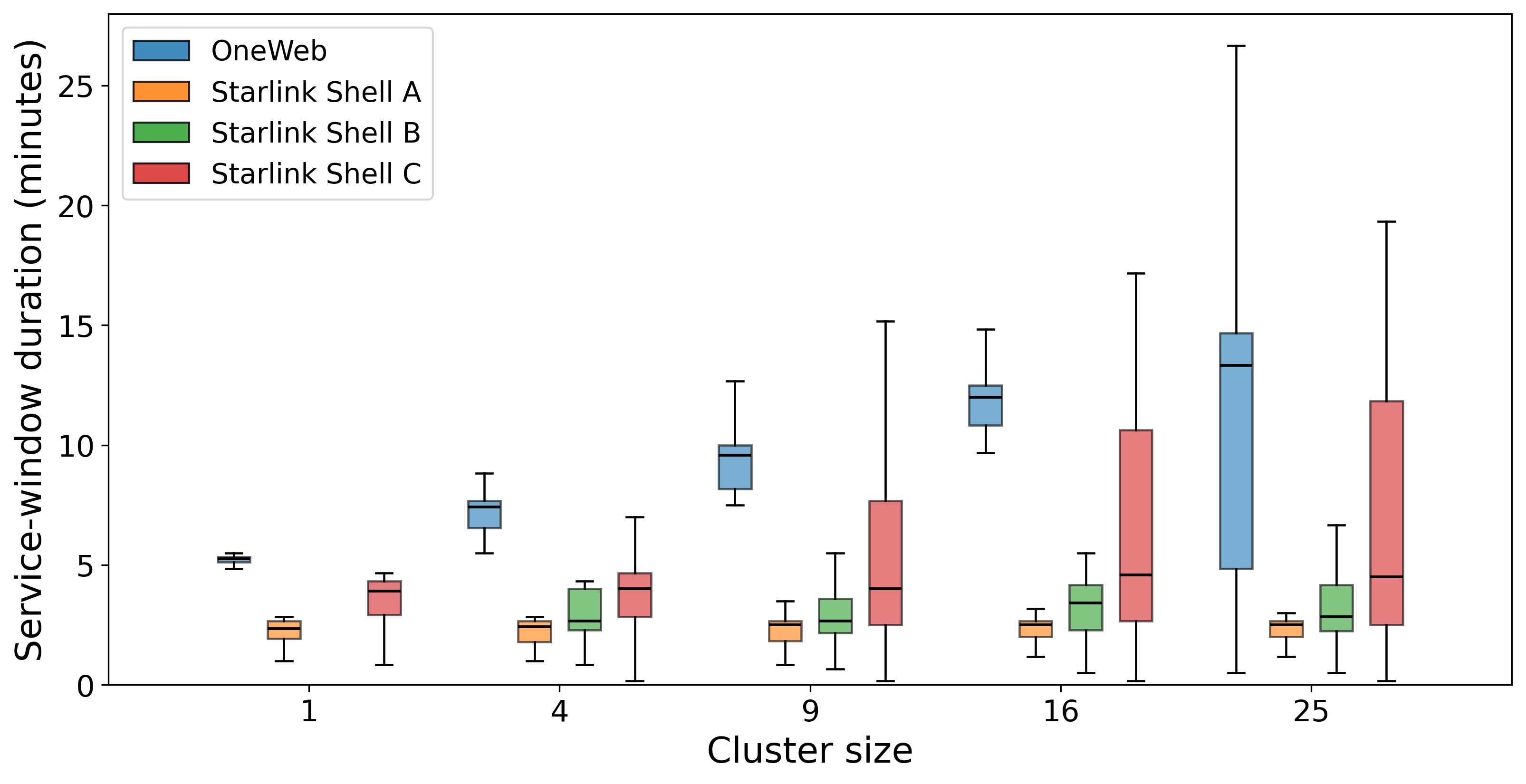}
        \caption{Polar.}
        \label{fig:service-window-polar}
    \end{subfigure}

    \caption{Service-window duration across regions and cluster sizes.}
    \label{fig:service-window}
    \vspace{-2mm}
\end{figure*}

\section{Problem Definition and Challenges}
\label{sec:problem}

\par \noindent\textbf{Cloud access is geographically uneven.}
Terrestrial cloud data centers can expose users to long access latency because intercontinental traffic largely depends on submarine fiber-optic cables, which carry over 99\% of international data traffic~\cite{liu2020scncriticality}. To quantify this limitation, we use the public cloud--edge latency dataset of Charyyev \emph{et al.}~\cite{charyyev2020latency}, which reports RTT measurements from RIPE Atlas probes to cloud locations. For each probe, we compute the nearest-cloud latency as the minimum RTT to any measured cloud data center and group the results by source region. As shown in Fig.~\ref{fig:nearest-cloud-latency}, Europe and North America achieve the lowest access latency, while South America and especially Africa experience much higher RTTs. This result shows that access to terrestrial cloud data centers remains geographically uneven even under an optimistic nearest-cloud assignment. This motivates SDCs, which enable satellite-generated data to be processed and disseminated cooperatively in orbit, closer to its source and with reduced dependence on geographically uneven terrestrial cloud access. 

\par \noindent\textbf{Single satellites provide only short service opportunities.}
Emerging space applications such as Earth observation and in-space AI generate large data volumes, including the $\sim$100~TB/day produced by NASA's SWOT and NISAR missions~\cite{nasa_swot_nisar_100tb}. However, downlink remains a major bottleneck: prior work reports that ground infrastructure can capture only about 2\% of available observations~\cite{denby2023kodan}. A single satellite is also not a practical service unit because its compute, storage, and connectivity are available only intermittently as it moves along its orbit. To illustrate this limitation, over representative equatorial, mid-latitude, and polar H3 cells \cite{uber_h3_repo}, together with Starlink gateways and OneWeb Satellite Network Portals (SNPs) \cite{zhao2025measuring}, shown in Fig.~\ref{fig:orbital-setup}, we conduct a two-line element (TLE)-driven analysis. Fig.~\ref{fig:service-window} shows that an individual satellite remains over a target H3 cell for only a few minutes, with typical median dwell times around 3--6~minutes across OneWeb and Starlink shells. Such short service windows are insufficient for persistent workloads including multi-stage processing, AI inference pipelines, caching, and replication, because unfinished tasks must be checkpointed, migrated, or delayed until another satellite becomes available.

\par
\begin{figure}[t]
    \centering
    \includegraphics[width=0.8\columnwidth]{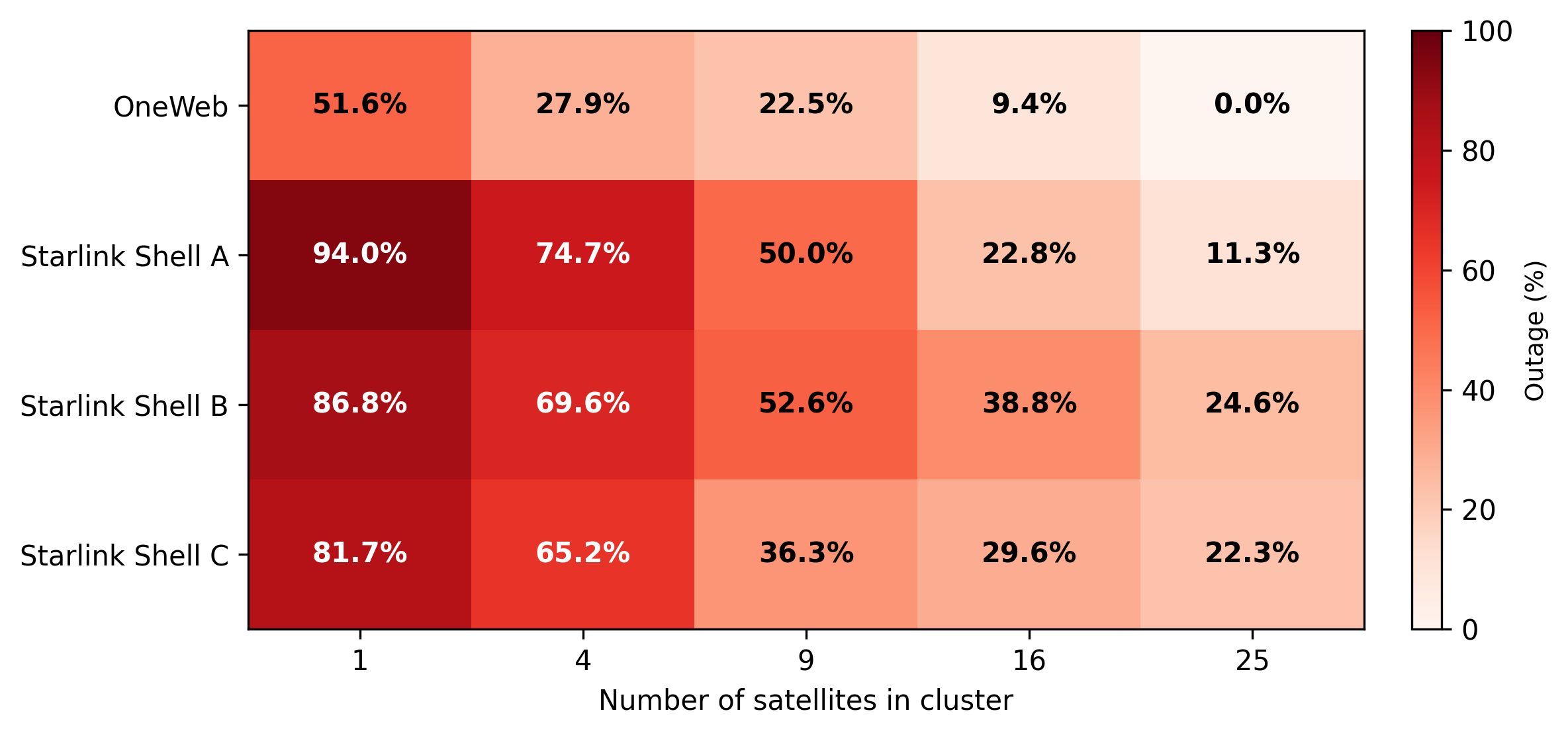}
    \caption{Direct gateway outage percentage versus SDC cluster size.}
    \label{fig:gateway-outage}
    \vspace{-2mm}
\end{figure}

 \noindent\textbf{Direct gateway access is also intermittent.} Even with sufficient onboard resources, a satellite may lack continuous direct gateway access for receiving inputs, coordinating with other satellites, or offloading results. Fig.~\ref{fig:gateway-outage} shows outages of 51.6\% for OneWeb and 94.0\% for Starlink Shell~A for a single satellite. Increasing the service-unit size reduces this outage, although the gain depends on the constellation shell and cluster size. Thus, a single satellite provides neither continuous regional service nor continuous direct gateway access. Grouping nearby satellites into a logical service unit mitigates both limitations: the cluster maintains longer service periods across equatorial, mid-latitude, and polar regions as members move through the target region (Fig.~\ref{fig:service-window}), while increasing the likelihood that at least one member has a direct gateway link (Fig.~\ref{fig:gateway-outage}). Therefore, clustering provides a practical SDC abstraction that aggregates compute, storage, and connectivity while reducing the impact of individual satellite mobility.

\par \noindent\textbf{Design implications.}
Realizing a moving satellite cluster as a practical SDC requires three key design decisions:
\begin{itemize}
    \item \textbf{Cluster size.} The SDC size must balance service continuity against coordination and forwarding overhead.

    \item \textbf{SDC architecture.} The system must define cluster-level addressing, membership management, border-satellite selection, and intra-SDC resource coordination.

    \item \textbf{Stateless multicast.} SDC workloads need efficient one-to-many delivery for replication, updates, AI parameter dissemination, and inter-SDC data placement without maintaining costly per-flow multicast state.
\end{itemize}

\par To address these requirements, the following sections present a topology-aware clustering analysis for selecting practical SDC sizes and a stateless multicast protocol for scalable one-to-many communication across dynamic satellite-based SDCs.

\section{LEO Satellite Cluster Size Selection}
\label{sec:leo-clustering}


\noindent\textbf{Network model.}
To capture LEO constellation dynamics, we divide time into discrete snapshots $\mathcal{T}=\{t_1,t_2,\ldots\}$. During each interval $(t_m,t_{m+1})$, the topology and active link states are assumed to be fixed. Because LEO satellite motion follows predictable orbital trajectories, the sequence of topology snapshots and the corresponding ISL states can be computed in advance from orbital information. Therefore, shortest-path trees, border-satellite selections, and BIER-DC forwarding assignments can be precomputed for upcoming snapshots and installed before scheduled topology transitions, reducing online recomputation and packet disruption. At snapshot $t$, the satellite network is modeled as an
undirected graph $\mathcal{G}^t=(\mathcal{S},\mathcal{E}^t)$, where
$\mathcal{S}$ is the set of satellites and $\mathcal{E}^t$ is the set of active laser ISLs. Ground gateways and user terminals are excluded from
the clustering objective and treated as part of the access layer, so the model captures the fundamental satellite-side coordination cost.

\noindent\textbf{SDC cluster characterization.}
At each snapshot $t$, the constellation is partitioned into $K$ logical SDC clusters $\mathcal{P}^t=\{S_{1,t},S_{2,t},\ldots,S_{K,t}\}$. For
cluster $S_{k,t}$, the induced intra-cluster ISL set is $E_{k,t}=\{(i,j)\in\mathcal{E}^t \mid i,j\in S_{k,t}\}$.
The shortest-path propagation delay between satellites $i$ and $j$ over the induced subgraph $(S_{k,t},E_{k,t})$ is denoted by $\delta_{k,t}(i,j)$. The intra-cluster communication cost is
\begin{equation}
C_{k,t}
=
\frac{1}{|S_{k,t}|(|S_{k,t}|-1)}
\sum_{\substack{i,j\in S_{k,t}\\i\neq j}}
\delta_{k,t}(i,j).
\label{eq:cluster-cost}
\end{equation}

A smaller $C_{k,t}$ indicates a more compact cluster with lower internal coordination delay, which is important because satellites within the same logical SDC are frequently involved in synchronization, replication, resource coordination, and intra-SDC delivery. Accordingly, the average intra-cluster shortest-path delay is used as a simple topology-derived proxy for coordination cost.

\noindent\textbf{Fixed-$K$ SDC cluster partitioning.}
For a given cluster count $K$, the goal is to compute a time-dependent
partition sequence $\mathcal{P}^t$ that minimizes the aggregate
intra-cluster communication cost across all snapshots.

\noindent\textbf{Input:} The dynamic constellation graph
$\mathcal{G}^t=(\mathcal{S},\mathcal{E}^t)$ for all
$t\in\mathcal{T}$, and the target cluster count $K$.

\noindent\textbf{Output:} A cluster partition
$\{S_{k,t}\}$ for all $k\in\{1,\ldots,K\}$ and
$t\in\mathcal{T}$.

\noindent\textbf{Goal:}
\begin{equation}
\min_{\{S_{k,t}\}}
\sum_{t\in\mathcal{T}}\sum_{k=1}^{K} C_{k,t}.
\label{eq:fixed-k-objective}
\end{equation}

\noindent\textbf{Constraints:}
\begin{align}
\bigcup_{k=1}^{K} S_{k,t} &= \mathcal{S},
\quad \forall t\in\mathcal{T}, \label{eq:partition-constraint}\\
S_{k,t}\cap S_{\ell,t} &= \emptyset,
\quad \forall k\neq \ell,\ \forall t\in\mathcal{T}, \label{eq:exclusion-constraint}\\
|S_{k,t}| &\geq 2,
\quad \forall k,\ \forall t\in\mathcal{T}, \label{eq:size-constraint}\\
(S_{k,t},E_{k,t}) &\text{ is connected},
\quad \forall k,\ \forall t\in\mathcal{T}. \label{eq:connectivity-constraint}
\end{align}
The first two constraints ensure that every satellite belongs to exactly
one SDC cluster. The size constraint avoids degenerate single-satellite
clusters, and the connectivity constraint ensures that all members of a
cluster can communicate through active intra-cluster ISLs.

\noindent\textbf{Two-stage parameter and scale selection.}
The fixed-$K$ objective is useful for comparing different assignments
under the same $K$, but it is not directly comparable across different
cluster counts. Therefore, we use the normalized pair-weighted
intra-cluster delay:
\begin{equation}
\bar{C}_{\mathrm{intra}}(K)
=
\frac{1}{|\mathcal{T}|}
\sum_{t\in\mathcal{T}}
\frac{
\sum_{k=1}^{K}|S_{k,t}|(|S_{k,t}|-1)C_{k,t}
}
{
\sum_{k=1}^{K}|S_{k,t}|(|S_{k,t}|-1)
}.
\label{eq:weighted-k-cost}
\end{equation}
We evaluate $\bar{C}_{\mathrm{intra}}(K)$ over candidate values of $K$
and select the elbow point $K^*$, where increasing the number of
clusters provides only marginal additional delay reduction. The
corresponding target cluster size is
$\bar{s}(K^*)=\frac{|\mathcal{S}|}{K^*}$.

\noindent\textbf{Solution methodology and scale selection.} Our fixed-$K$ formulation is closely related to Min-Sum $k$-Clustering,
an NP-hard problem that partitions vertices under graph-induced distances
to minimize intra-cluster pairwise communication cost~\cite{naderi2024approximation}.
Therefore, we use a scalable heuristic to construct compact and connected SDC clusters. For each candidate $K$, we select seed satellites and assign each remaining satellite to its nearest seed based on Dijkstra shortest-path distances over the weighted ISL graph. We then keep only assignments whose induced cluster subgraphs are connected. The resulting partitions are evaluated using the normalized pair-weighted intra-cluster delay, and the elbow point $K^*$ is selected as the point where additional clusters provide only marginal delay reduction. This defines the target average cluster size $\bar{s}(K^*)=|\mathcal{S}|/K^*$.

\par The target size $\bar{s}(K^*)$ specifies how many satellites should form an SDC, but not how these satellites should be arranged in the orbital topology. We therefore represent each practical SDC as a rectangular $p \times q$ block, consisting of $p$ adjacent orbital planes moving in the same direction and $q$ consecutive satellites in each plane, for a total of $pq$ satellites. This co-moving structure limits relative motion and helps maintain stable relative positions over time. We evaluate candidate blocks whose size $pq$ is close to $\bar{s}(K^*)$ by placing each block at different positions in the constellation and examining it over all topology snapshots. Placements that do not induce a connected intra-SDC subgraph are discarded. Among the remaining candidates, we select the compact block that is closest to the target size while providing low average intra-SDC shortest-path delay and small diameter.

\section{BIER-DC Design}
\label{sec:bier-dc-design}
\subsection{BIER-DC Overview}

\begin{figure}[t]
    \centering
    \includegraphics[width=\columnwidth]{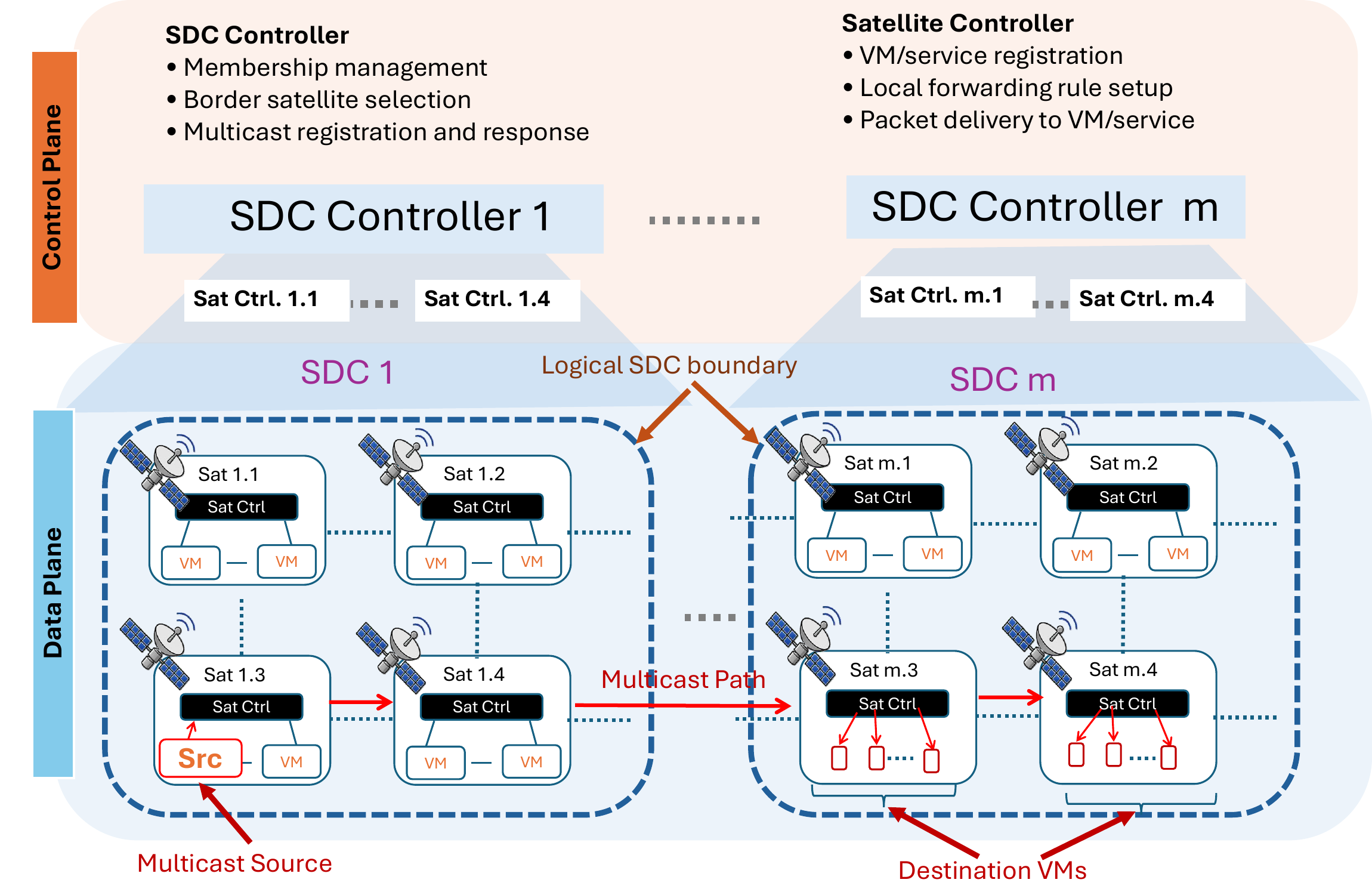}
    \caption{BIER-DC architecture with hierarchical SDC control and cluster-level data forwarding.}
    \label{fig:bierdc-architecture}

    \vspace{-10pt}
    
\end{figure}

We present BIER-DC, a stateless multicast architecture for in-orbit space data centers. As shown in Fig.~\ref{fig:bierdc-architecture}, BIER-DC organizes satellites into logical space data center clusters, called SDCs, and separates multicast forwarding into two levels: inter-SDC delivery across clusters and intra-SDC delivery inside each cluster.

\noindent{\bf 1) Design principles:} The design of BIER-DC follows three primary goals. {\it First, compact multicast headers.} Since satellite links have limited bandwidth and energy budget, BIER-DC should avoid per-receiver or per-VM bit positions in the inter-SDC header. {\it Second, scalable stateless forwarding.} Multicast forwarding should not require per-flow forwarding state at every satellite, especially as the number of destination satellites, VMs, containers, and multicast groups increases. {\it Third, hierarchical control.} The control plane should match the clustered structure of space data centers: SDC controllers coordinate cluster-level forwarding, while satellite controllers handle local resource management and final service delivery.

\noindent{\bf 2) Key ideas of BIER-DC:} BIER-DC adopts three key ideas to realize these goals: {\bf (i) cluster-level addressing.} Satellites are grouped into SDCs, denoted as SDC~1 to SDC~$m$, and each SDC is assigned a unique \emph{Cluster ID}. The Cluster ID is used in control messages, forwarding entries, and BIER-DC packet headers. We use $m$ to denote the number of SDCs and $n$ to denote the number of satellites per SDC. This abstraction allows BIER-DC to describe multicast delivery at the SDC level rather than at the individual satellite, VM, or container level, thereby reducing header growth and limiting forwarding-state complexity as the number of service endpoints increases.  {\bf (ii) border-satellite-based inter-SDC forwarding.} Instead of encoding every destination satellite or VM in the packet header, BIER-DC encodes compact forwarding entries that identify the destination SDC and the selected border satellite. This allows the source SDC to express the inter-SDC multicast tree using a small number of cluster-level entries. {\bf (iii) local header translation at the destination SDC.} When a packet reaches a border satellite of a destination SDC, the inter-SDC BIER-DC header is replaced by a local intra-SDC header. The destination SDC then delivers the packet to the selected satellite, VM, container, or data object using local forwarding decisions. 

\noindent\textbf{3) System overview and workflow:} 
The BIER-DC architecture consists of a \textit{data plane} and a \textit{control plane}. In the data plane, satellites inside each SDC are connected through intra-cluster ISLs, while neighboring SDCs are connected through inter-cluster ISLs. Each satellite may host local computation and storage resources, represented as VMs or containers, enabling in-orbit services such as data replication, caching, processing, and multicast delivery.
 
The control plane follows a hierarchical structure. Each satellite has a local satellite controller, which abstracts functions already present in modern onboard systems~\cite{nasa_cdhs,nasa_satcom}. BIER-DC extends this role to networking and service management: the satellite controller monitors neighboring ISLs, applies forwarding rules, tracks local resources, manages VM/container placement when onboard edge computing is available, and performs final delivery to the selected VM, container, or data object.

Each SDC is managed by an SDC controller that maintains control sessions with the satellite controllers in its cluster. The SDC controller keeps cluster-level state, including the Cluster ID, SDC membership, intra-cluster topology, registered border satellites, aggregate per-satellite capacity, and forwarding entries for intra-SDC and inter-SDC delivery. This design mirrors cloud and edge control planes, where a node-level agent manages local workloads and reports status to a higher-level controller. For example, Kubernetes uses kubelet as the node agent, and KubeEdge extends Kubernetes-native orchestration to edge hosts, including a cloud-native satellite case study~\cite{kubernetes_nodes, kubernetes_kubelet,kubeedge,kubeedge_satellite}. Accordingly, BIER-DC keeps only an aggregate view at the SDC controller, while fine-grained placement and final delivery decisions remain with satellite controllers. 
 
\textit{\bf Border satellites} are the managed entry and exit points of each SDC. They are edge satellites that maintain ISL connections to neighboring SDCs. Each border satellite registers with its local SDC controller, which records its satellite ID, boundary face, and reachable neighboring SDCs. Only registered border satellites are used for inter-SDC forwarding; inter-SDC edges incident to non-registered satellites are excluded from the forwarding graph. On egress, a border satellite forwards cloned packets toward the selected neighboring SDC. On ingress, it hands incoming packets to the local intra-SDC delivery process. If multiple border satellites can reach the same neighboring SDC, BIER-DC selects the lowest-cost candidate. SDC members, including border satellites, are selected from the same or adjacent orbital planes and move in the same direction, so their relative positions in the topology remain approximately stable over extended periods. This stability allows registered border satellites to remain active without frequent reselection. Moreover, predictable orbital motion enables inter-SDC border connectivity and scheduled handovers to be precomputed for upcoming topology snapshots.

\subsection{BIER-DC Header and Addressing}

\par BIER-DC first requires identifying where a multicast packet should be delivered and how it should cross the space data center topology. Unlike conventional satellite-level multicast encoding, which represents individual satellites or egress nodes in the packet header, BIER-DC uses a cluster-level addressing model. Each SDC is assigned a unique \emph{Cluster ID}, denoted by $\mathrm{CID}_k$, which serves as the basic inter-SDC forwarding identifier. Each satellite also has a local \emph{Satellite ID}, which is used for intra-SDC delivery, border-satellite registration, and ingress-satellite resolution. Services hosted inside an SDC, such as VMs, containers, storage replicas, or application instances, are identified by a local service or VM identifier. This two-level addressing model keeps inter-SDC forwarding compact while still allowing precise delivery inside each destination SDC.

\par \emph{\bf Addressing scheme.} BIER-DC follows a four-step process to address destinations. \emph{\bf Step-I: assigning SDC identifiers.} Each SDC is assigned a unique cluster identifier and packets are forwarded across the SDC-level topology using these identifiers, instead of carrying a flat satellite-level multicast bitstring. \\ \emph{\bf Step-II: identifying border satellites.} Each SDC maintains a local list of border satellites that connect the SDC to neighboring SDCs. These satellites act as ingress or egress points for inter-SDC forwarding and can be represented by compact local indexes. 
\\ \emph{\bf Step-III: identifying internal destinations.} Inside each SDC, satellites and hosted services are assigned local identifiers. Therefore, a complete service address can be written as $[\mathrm{SDC}, \mathrm{Border}, \mathrm{Satellite}, \mathrm{VM}]$, where the SDC and border fields guide inter-SDC forwarding, while the satellite and VM fields are used for local delivery after the packet reaches the destination SDC. 
\\ \emph{\bf Step-IV: maintaining the address mapping.} The SDC controller maintains the mapping between destination services, internal satellites, and selected border satellites. When the topology or SDC membership changes, these mappings are updated and forwarding assignments are re-evaluated over the updated topology snapshot.

\begin{figure}[!t]
    \centering
    \includegraphics[width=\columnwidth,height=0.22\textheight,keepaspectratio]{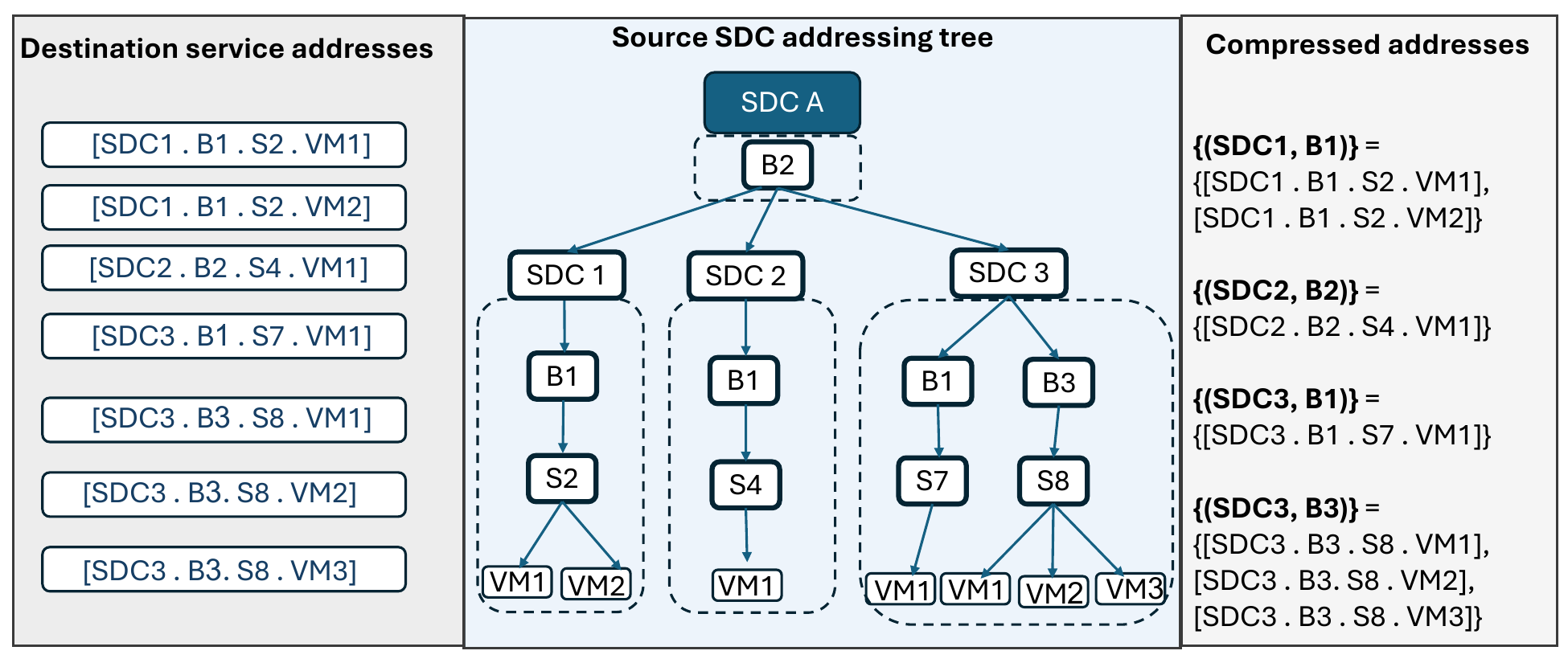}
    \caption{BIER-DC addressing example. Destination service addresses are compressed into compact SDC--border-satellite entries for inter-SDC forwarding.}
    \label{fig:bierdc-addressing-example}
    
    \vspace{-15pt}
\end{figure}

\par Fig.~\ref{fig:bierdc-addressing-example} shows an example of the BIER-DC addressing model. Each destination service is first represented by a hierarchical address in the form $[\mathrm{SDC}, \mathrm{Border}, \mathrm{Satellite}, \mathrm{VM}]$. The SDC and border-satellite fields are used for inter-SDC forwarding, while the satellite and VM fields are interpreted only inside the destination SDC. As shown in the figure, multiple destination services may share the same SDC and selected border satellite. For example, two services in SDC~1 behind border satellite $B_1$ can be compressed into the same inter-SDC entry $(\mathrm{SDC1},B_1)$. Similarly, multiple services in SDC~3 can be represented by compact border-satellite entries depending on the selected ingress border. This aggregation allows BIER-DC to avoid inserting all service-level destinations into the inter-SDC packet header.

\par \emph{\bf Packing compressed addresses into packets.}
To indicate multicast forwarding, the source SDC packs a compressed destination representation into the BIER-DC header. Instead of inserting every service address into each packet, BIER-DC aggregates services that share the same destination SDC and selected border satellite into one $(\mathrm{SDC},\mathrm{Border})$ entry. The destination SDC later expands this entry into the corresponding local satellite and service destinations using its internal mapping table.

\par The destination path encoding is the main scalability feature of BIER-DC. Let $m$ be the number of SDCs. For an SDC-level path $[\mathrm{CID}_{k_1}, \mathrm{CID}_{k_2}, \ldots, \mathrm{CID}_{k_h}]$, the source SDC $\mathrm{CID}_{k_1}$ is implicit, so only the remaining $h-1$ SDC hops are encoded. Each hop identifies the next SDC and the selected ingress border satellite. Under fixed-width encoding, each entry uses an SDC index of $\lceil \log_2 m \rceil$ bits and an ingress-border-satellite index of $\lceil \log_2 n_2 \rceil$ bits, where $n_2$ is the maximum number of border satellites on the longest SDC face in the current topology snapshot. Thus,
\begin{equation}
w_{\mathrm{entry}}
=
\lceil \log_2 m \rceil
+
\lceil \log_2 n_2 \rceil
\quad \text{bits}.
\end{equation}
The ingress-border-satellite index is interpreted locally within the destination SDC: the selected inter-SDC crossing determines the ingress face, and the encoded index selects a border satellite from the sorted list on that face. In our simulation, this gateway identity is read from the precomputed shortest-path tree state. In deployment, the destination SDC can reconstruct it from its local face-to-border-satellite map, making the header self-contained without a flat global satellite identifier.

\begin{figure}[!t]
    \centering
\includegraphics[width=\columnwidth,height=0.18\textheight]{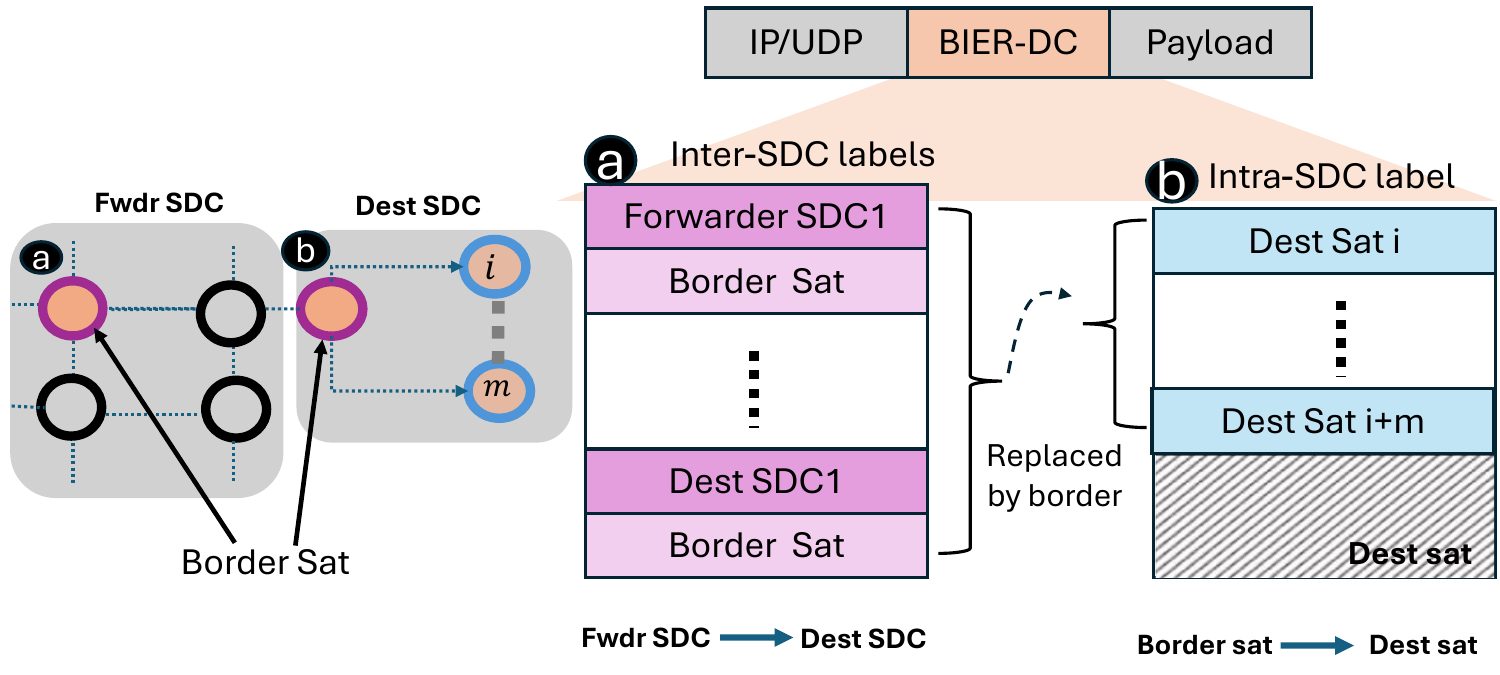}
    \caption{The inter-SDC label used by satellite $a$ in the forwarding SDC is replaced by an intra-SDC label at ingress satellite $b$ in the destination SDC.}
    \label{fig:bierdc-header}
    \vspace{-15pt}
\end{figure}

\par Fig.~\ref{fig:bierdc-header} illustrates the packet-header operation. A packet carries an IP/UDP header, followed by the BIER-DC header and payload. During inter-SDC forwarding, the BIER-DC header carries labels for the current forwarding SDC, destination SDC, and selected destination border satellite. Once the packet reaches the selected border satellite of a destination SDC, the inter-SDC label is replaced by an intra-SDC label that identifies the internal destination satellite or service group. This allows the same packet format to support inter-SDC routing and local intra-SDC delivery while keeping the wide-area forwarding header compact.

\par For a destination path of $h$ SDCs, the encoding length is $(h-1)w_{\mathrm{entry}}$ bits. For multicast traffic, the source concatenates the path encodings required by the destination set. Intermediate SDCs group destinations that share the same next SDC and ingress border satellite, so one forwarded packet copy can serve multiple downstream destinations. If destinations require different ingress satellites into the same next SDC, BIER-DC creates separate entries and packet copies for those branches, preserving the selected SDC-level distribution tree without a flat per-satellite multicast bitstring.

\par Unlike satellite-level encoding, which scales as $O(|\mathcal{S}|)$, BIER-DC reduces overhead by scaling with the number of SDC-level hops. For a destination set $\mathcal{D}$, the overhead is proportional to
\[
O\!\left(
\sum_{d\in\mathcal{D}} (h_d-1)
\left(
\lceil \log_2 m\rceil + \lceil \log_2 n_2\rceil
\right)
\right).
\]
Here, $h_d$ is the number of SDCs on the path to destination $d$, $m$ is the number of SDCs, and $n_2$ is the maximum number of border satellites on the longest SDC face. Since $m \ll |\mathcal{S}|$, BIER-DC replaces flat satellite-level bitstrings with two-level addressing: compact path entries route packets through SDCs, while local labels deliver packets to final satellites, VMs, containers, or services inside each destination SDC. This hierarchy lets the SDC controller manage aggregate inter-SDC mappings and the local satellite controller handle final delivery. Forwarding assignments are re-evaluated whenever the topology or SDC membership changes over the updated topology snapshot.

\subsection{Forwarding Mechanism}

\par BIER-DC forwards multicast packets at two levels: inter-SDC forwarding between space data centers and intra-SDC delivery inside destination SDCs. Inter-SDC forwarding is performed by border satellites, which process the compact path encoding carried in the BIER-DC header. Each remaining destination is associated with an SDC-level path and a selected ingress border satellite for each downstream SDC. When a packet reaches a border satellite, the satellite determines whether the current SDC contains local destinations and whether the packet must continue toward one or more adjacent SDCs.

\par Algorithm~\ref{alg:bierdc-border-forwarding} gives the border-node forwarding procedure. The algorithm takes as input a packet $p$ with BIER-DC header $H$, the current SDC $c$, the ingress border satellite $b_{\mathrm{in}}$, and the local SDC controller $\Gamma_c$. It first extracts the set of unresolved destinations $\mathcal{D}_{loc}$ that belong to the current SDC and groups the
remaining destinations by their next adjacent SDC into $\mathcal{G}_{down}$. This grouping allows BIER-DC to send only one packet copy for destinations that share the same next SDC and to replicate packets only when the downstream paths diverge.

\par \textbf{Case 1: Local delivery inside the current SDC.}
If the current SDC contains one or more destinations, the ingress border satellite queries the local SDC controller to resolve the corresponding destination satellites, services, or VMs. The controller then constructs an intra-SDC header rooted at $b_{\mathrm{in}}$. This local header replaces the inter-SDC BIER-DC header only for the packet copy delivered inside the current SDC. Therefore, local delivery can be handled using compact intra-SDC forwarding information without exposing satellite-level destinations in the inter-SDC header.

\par \textbf{Case 1.1: A single local destination satellite.}
If the local destination set contains only one destination satellite, no local multicast branching is needed. The border satellite creates a local packet copy, replaces its BIER-DC header with the intra-SDC header, and forwards the packet along the selected local path toward that destination satellite.

\par \textbf{Case 1.2: Multiple local destination satellites.}
If the current SDC contains multiple local destination satellites, the controller identifies the required local branches from the ingress border satellite. The border satellite then creates one packet copy per local branch and attaches the corresponding intra-SDC header to each copy. Packets are replicated only at the required branching points, which avoids carrying a flat satellite-level bitstring in the inter-SDC header.

\par \textbf{Case 2: Inter-SDC forwarding to downstream adjacent SDCs.}
After local destinations are processed, the border satellite forwards the packet toward unresolved destinations in downstream SDCs. For each group in $\mathcal{G}_{down}$, the border satellite selects the egress border satellite toward the next adjacent SDC and forwards the packet to it using the shortest path inside the current SDC. The egress border satellite then transmits the packet over the inter-SDC link to the selected ingress border satellite in the next SDC. Before forwarding, the packet header is updated to keep only the destinations belonging to that downstream branch, the current SDC entry is consumed, and the TTL is decremented. If the packet has no local destinations and no downstream SDCs, it is dropped.

\par This design separates inter-SDC forwarding from intra-SDC delivery. The inter-SDC header only carries compact SDC-level path information, while detailed satellite-level delivery is resolved locally inside each destination SDC. Whenever the topology or SDC membership changes, the forwarding assignment should be re-evaluated over the updated topology snapshot.

\begin{algorithm}[t]
\caption{BIER-DC Border-Node Forwarding}
\label{alg:bierdc-border-forwarding}
\begin{algorithmic}[1]
\Require Packet $p$ with BIER-DC header $H$, current SDC $c$,
         ingress border satellite $b_{\mathrm{in}}$, SDC controller $\Gamma_c$
\Ensure Forwarding decision for $p$

\State $\mathcal{D}_{loc} \gets \mathrm{LocalDestinations}(H,c)$
\State $\mathcal{G}_{down} \gets \mathrm{GroupByNextAdjacentSDC}(H,c)$

\PinkCase{Case 1: Local delivery inside the current SDC.}
\If{$\mathcal{D}_{loc}\neq\emptyset$}
    \State $\mathcal{L} \gets \Gamma_c.\mathrm{Resolve}(\mathcal{D}_{loc})$
    \State $H_{intra} \gets \Gamma_c.\mathrm{BuildIntraHeader}(b_{\mathrm{in}},\mathcal{L})$

    \PinkCase{Case 1.1: A single local destination satellite.}
    \If{$|\mathcal{L}|=1$}
        \State $p_l \gets \mathrm{CloneAndReplaceHeader}(p,H_{intra})$
        \State Forward $p_l$ to the only destination satellite in $\mathcal{L}$

    \PinkCase{Case 1.2: Multiple local destination satellites.}
    \Else
        \ForAll{$q \in \mathrm{LocalBranches}(b_{\mathrm{in}},\mathcal{L})$}
            \State $p_q \gets \mathrm{CloneAndReplaceHeader}(p,H_{intra}(q))$
            \State Forward $p_q$ toward local branch $q$
        \EndFor
    \EndIf
\EndIf

\PinkCase{Case 2: Inter-SDC forwarding to downstream SDCs.}
\ForAll{$g \in \mathcal{G}_{down}$}
    \State $(c_{next},\mathcal{D}_g) \gets g$
    \State $b_{\mathrm{out}} \gets \mathrm{SelectBorder}(c,c_{next})$
    \State $P \gets \mathrm{ShortestPath}(b_{\mathrm{in}},b_{\mathrm{out}})$
    \State $p_g \gets \mathrm{Clone}(p)$
    \State Update $p_g.H$ to keep only $\mathcal{D}_g$
    \State Consume the current SDC entry in $p_g.H$
    \State Decrement $p_g.H.\mathrm{TTL}$
    \State Forward $p_g$ along $P$ and transmit it to $c_{next}$
\EndFor

\If{$\mathcal{D}_{loc}=\emptyset$ and $\mathcal{G}_{down}=\emptyset$}
    \State Drop $p$
\EndIf

\end{algorithmic}
\end{algorithm}

\noindent{\bf Lemma 1 (Shortest-Path Concatenation).} \label{lem:shortest-path-concat}
Consider snapshot $\mathcal{G}^t=(\mathcal{S},\mathcal{E}^t)$ with
shortest-path delay $\delta^t(i,j)$. For an ordered sequence
$v_0,\ldots,v_k \in \mathcal{S}$, let $P^t_{v_{i-1},v_i}$ denote a
shortest path between $v_{i-1}$ and $v_i$. The concatenation
$P^t_{v_0,v_1}\circ\cdots\circ P^t_{v_{k-1},v_k}$ is a shortest path
from $v_0$ to $v_k$ if and only if
\[
    \delta^t(v_0,v_k)=\sum_{i=1}^{k}\delta^t(v_{i-1},v_i).
\]

\noindent{\it Proof sketch.}
The concatenated cost equals $\sum_{i=1}^{k}\delta^t(v_{i-1},v_i)$ since
each segment is shortest. If this sum equals $\delta^t(v_0,v_k)$, the
route achieves the minimum possible delay and is therefore shortest. The
converse holds directly from the definition of $\delta^t$.
\hfill$\square$

By Lemma~1, BIER-DC preserves shortest-path routing because it does not recompute independent local routes between SDC borders. Instead, each border-to-border segment is extracted from the original shortest-path tree computed for the current topology snapshot. Therefore, the selected SDC border satellites appear in the same order as they do on the original SPT path from the source ingress satellite to the destination. Concatenating these extracted segments reconstructs the same end-to-end shortest route. Hence, BIER-DC represents the original SPT result at the SDC-border level without loss of routing optimality.

\par {\bf Handling Failures in BIER-DC.} BIER-DC handles satellite and ISL failures through snapshot-based recomputation and localized recovery. Because the packet header encodes SDC-border-level decisions derived from the current shortest-path tree, a failed satellite or inter-SDC link may invalidate one or more border-to-border segments. Upon detection, the affected satellite or SDC controller updates its connectivity view, reports the change to the SDC-level control plane, and recomputes the affected SPT-derived segments for the next topology snapshot. During this transition, packets with stale border-level information are either locally detoured toward the same next border satellite if an alternative path exists, or dropped and recovered by upper-layer reliability or source-side retransmission. Because the LEO topology evolves over time, the assignment should be re-evaluated over the updated topology snapshot.

\section{Evaluation}
\label{sec:evaluation}
\par Evaluations in this section focus on answering the following three questions related to \textsc{BIER-DC}.  \textbf{Q1:} Can \textsc{BIER-DC} achieve high multicast goodput for data delivery compared with existing multicast forwarding approaches? \textbf{Q2:} Can \textsc{BIER-DC} maintain compact packet headers as the number of VMs or services inside each SDC increases? \textbf{Q3:} How does \textsc{BIER-DC} scale to larger space data center deployments with an increasing number of SDCs?

\textbf{Prototype and testbed setup.}  We evaluate \textsc{BIER-DC} using NetSatBench~\cite{detti2026netsatbench}, a container-based LEO emulator that instantiates satellites as independent Docker containers and connects them according to time-varying constellation snapshots. The snapshots are generated using StarPerf~\cite{lai2020starperf}, which provides satellite positions, ISL topology, and propagation delays for the target constellation. In our prototype, each satellite container additionally runs a lightweight \textsc{BIER-DC} runtime agent that parses \textsc{BIER-DC} headers and forwards multicast packets either to the next hop or to a local VM/service. A centralized controller computes forwarding rules based on the source, destinations, SDC membership, and inter-SDC paths, and installs them through the emulator control channel. For each experiment, a UDP sender in the source container segments the input file according to the available payload after subtracting the protocol header size, while destination receivers record delivery statistics. By varying the number of VMs per satellite and the number of SDCs, we measure header size, raw payload size, packet count, and goodput. 

\textbf{Compared protocols.} We compare \textsc{BIER-DC} with BIER~\cite{rfc8279}, BIER-TE~\cite{rfc9262}, and Yeti~\cite{diab2022yeti}, since these stateless or packet-carried multicast schemes can be evaluated over the same LEO topology, source, and destination set. This enables a direct comparison of multicast header overhead and its effects on payload efficiency and goodput. We exclude terrestrial data center schemes such as Elmo~\cite{shahbaz2019elmo} and Orca~\cite{diab2022orca} because they assume fixed Clos/leaf-spine fabrics, stable rack structure, and terrestrial controller/switch abstractions. Adapting them to dynamic inter-satellite links and changing SDC membership would require architectural redesign, making the comparison implementation-dependent.

\begin{figure}[t]
    \centering
    \includegraphics[width=0.7\columnwidth]{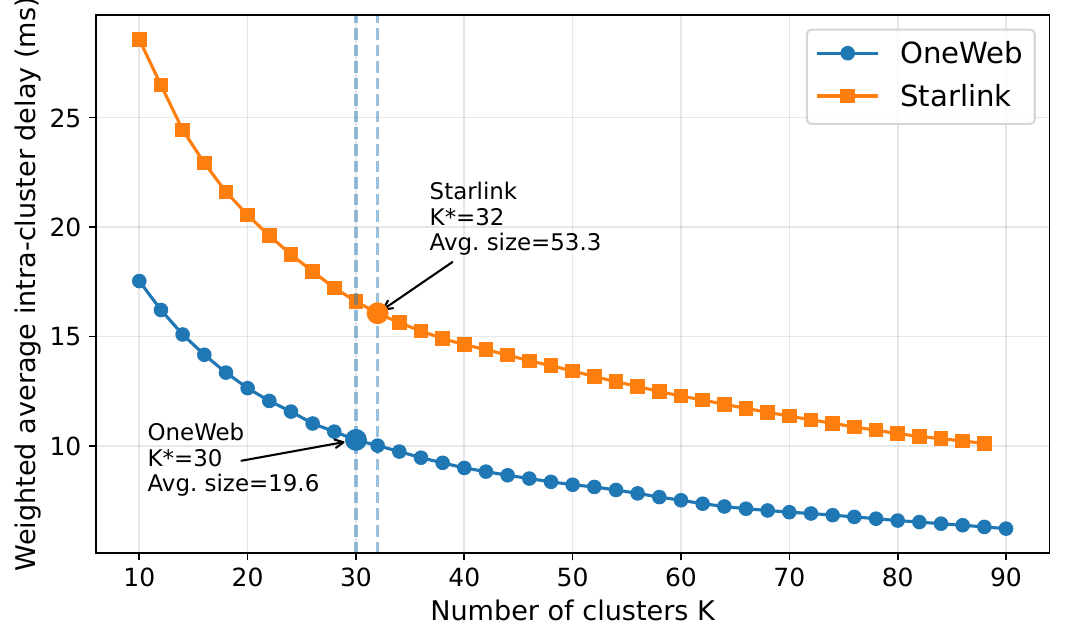}
    \caption{Elbow-based cluster-count selection for OneWeb and Starlink.}
    \label{fig:cluster-count-selection}
    \vspace{-15pt}
\end{figure}

\begin{figure*}[t]
    \centering
    \makebox[\textwidth][c]{%
        \subfloat[VM scalability.]{
            \includegraphics[width=0.30\textwidth]{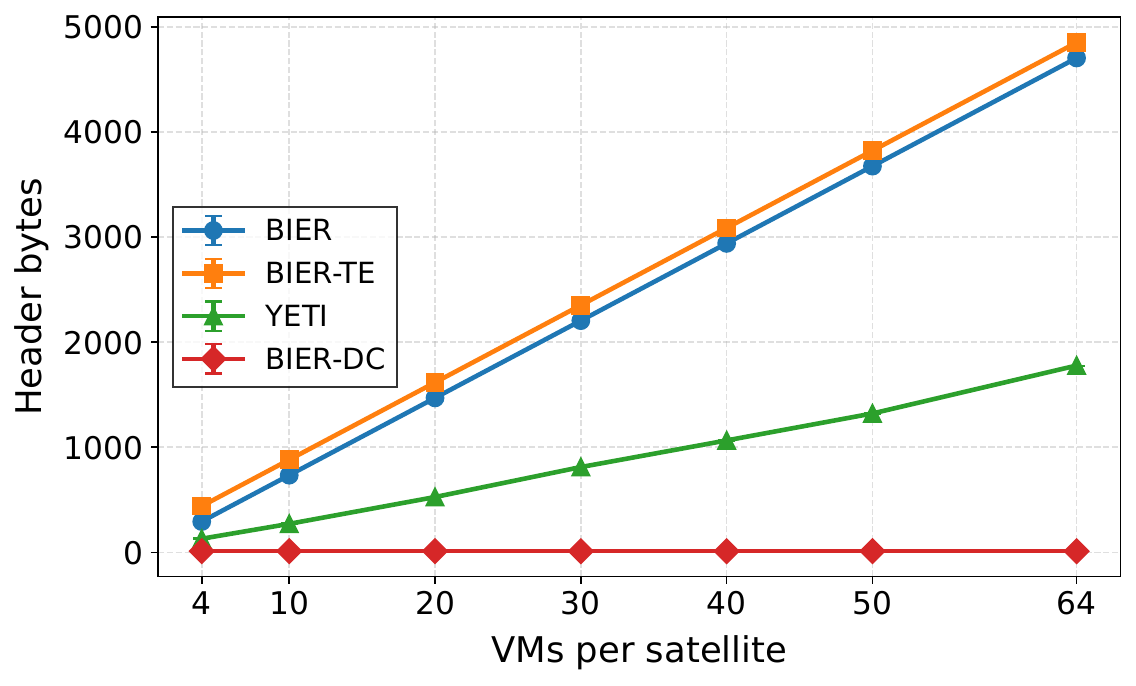}
            \label{fig:vm-scalability}
        }
        \hspace{0.02\textwidth}
        \subfloat[Destination satellite density.]{
            \includegraphics[width=0.30\textwidth]{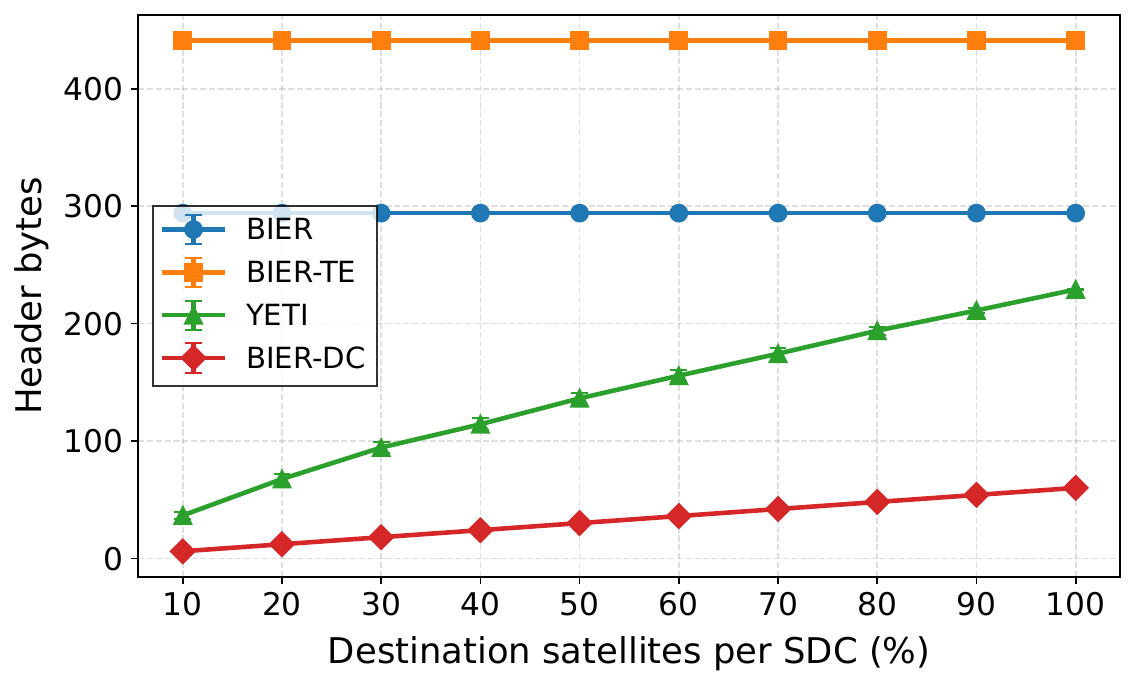}
            \label{fig:destination-density}
        }
        \hspace{0.02\textwidth}
        \subfloat[Destination SDC percentage.]{
            \includegraphics[width=0.30\textwidth]{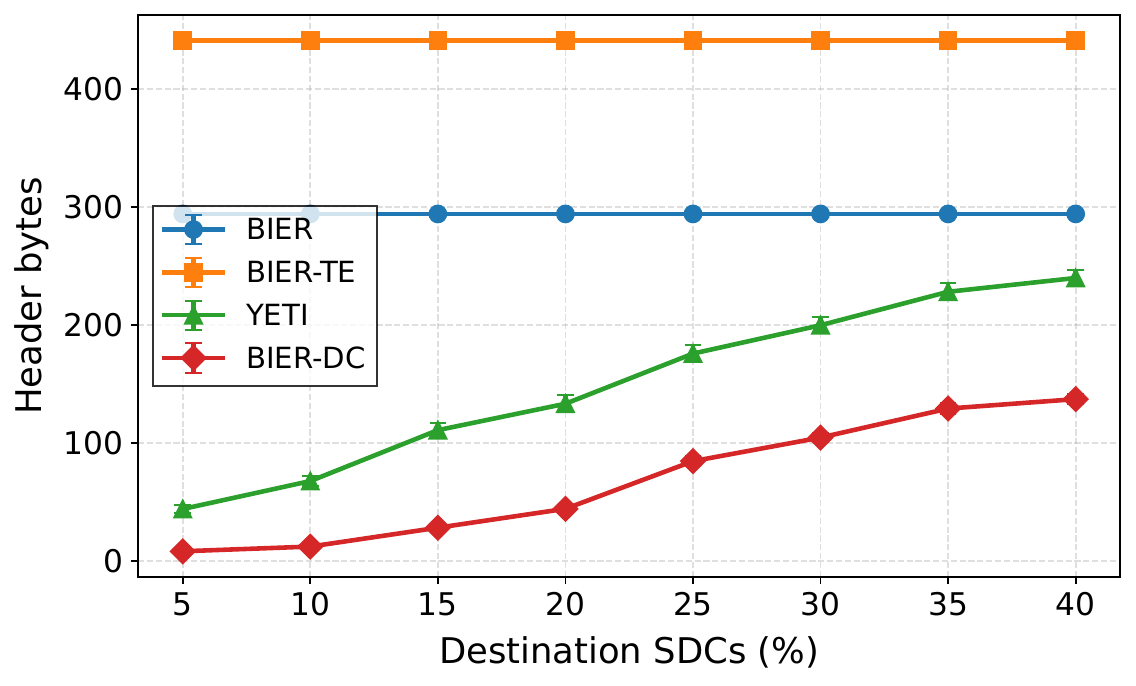}
            \label{fig:destination-sdc-scaling}
        }
    }

    \caption{Header overhead comparison in the OneWeb topology across three evaluation scenarios: (a) VM scalability, (b) destination satellite density, and (c) destination SDC scaling.}
    \label{fig:header-overhead-scenarios}
    \vspace{-2mm}
\end{figure*}

\textbf{Constellation, SDC, workload, and multicast setting.}
We evaluate \textsc{BIER-DC} on OneWeb with 588 satellites and a Starlink shell with 1705 satellites in a $+$grid topology. Satellites are grouped into SDCs that act as logical clusters for computing, storage, and multicast services. In each experiment, one source satellite multicasts an EO data object to destination VMs/services across one or more SDCs. Each endpoint is assigned a globally unique address, $\langle \mathrm{SDC\ ID}, \mathrm{Satellite\ ID}, \mathrm{VM/Service\ ID}\rangle$, while \textsc{BIER-DC} keeps the inter-SDC header compact by encoding paths at the SDC level. We vary the number of virtualized service endpoints per satellite from $M=4$ to $M=64$. The lower bound, $M=4$, represents a conservative core-pinned setting consistent with legacy radiation-hardened processors such as the four-core GR740~\cite{gr740}. Higher densities reflect heterogeneous satellite-edge platforms such as Jetson AGX Orin, with up to 12 Arm Cortex-A78AE CPU cores, 2,048 CUDA cores, and 64 Tensor Cores~\cite{nvidiaJetsonAGXOrinSpec}. Thus, $M=16\text{--}32$ represents accelerator-assisted satellite-edge payloads, while $M=64$ is a dense future SDC stress point for multicast header scaling. For each VM-scale and SDC-scale configuration, the same input file is transferred over UDP from the source container to all receivers, with packet payload adjusted to each protocol's header size. Moreover, Fig.~\ref{fig:cluster-count-selection} shows that increasing the number of SDCs reduces
the weighted average intra-cluster delay, but the reduction becomes marginal
after $K^*=30$ for OneWeb and $K^*=32$ for Starlink. These elbows correspond
to target average SDC sizes of 19.6 and 53.3 satellites, respectively.
Accordingly, we select a $4\times5$ OneWeb footprint, which almost exactly
matches the target size, and a compact $7\times7$ Starlink footprint, which
remains close to the target while maintaining a balanced shape and small
intra-SDC diameter.

\par \textbf{Header overhead under SDC-scale multicast.}  Fig.~\ref{fig:header-overhead-scenarios} compares the packet-header overhead of \textsc{BIER-DC}, Yeti, BIER, and BIER-TE under three evaluation scenarios. In all experiments, we select one source satellite inside one SDC and multicast traffic toward destination satellites located in multiple destination SDCs. To capture the effect of SDC-level path diversity, we select destination satellites distributed across SDCs located one, two, and three hops away from the source SDC. This setup allows us to evaluate whether each protocol can maintain compact packet headers when the multicast group expands in three dimensions: the number of VMs per satellite, the density of destination satellites inside each selected SDC, and the number of destination SDCs involved in the multicast session.

\par
\begin{figure}[t]
    \centering
    \includegraphics[width=0.60\columnwidth]
    {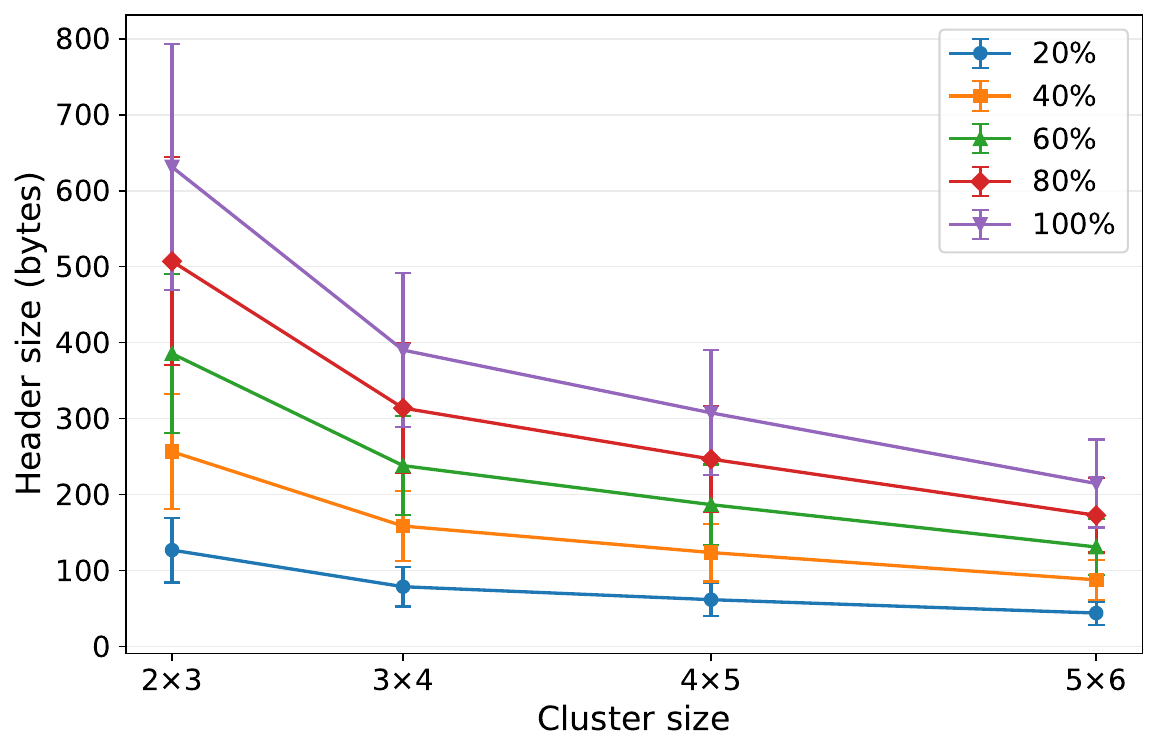}
    \caption{BIER-DC header size versus SDC cluster size.}
    \label{fig:cluster_size_sensitivity}
    \vspace{-15pt}
\end{figure}

 As shown in Fig.~\ref{fig:vm-scalability}, the header overhead of BIER and BIER-TE increases rapidly as the number of VMs per satellite grows, because their bitstring-based encoding must reserve space for a larger forwarding or adjacency space. Yeti reduces this overhead compared with BIER and BIER-TE, but its header still grows with the number of service endpoints. In contrast, \textsc{BIER-DC} keeps the header nearly constant because it encodes forwarding at the SDC level and uses compact SDC/ingress-border identifiers instead of per-VM or per-satellite bit positions. Fig.~\ref{fig:destination-density} further shows that increasing the percentage of destination satellites inside each selected SDC has only a moderate impact on \textsc{BIER-DC}, while Yeti grows more noticeably. Finally, Fig.~\ref{fig:destination-sdc-scaling} shows that when the multicast session spans more destination SDCs, \textsc{BIER-DC} grows gradually with the number of encoded SDC-level paths, but remains substantially smaller than the bitmap-based baselines. These results confirm that \textsc{BIER-DC} is better suited for SDC-based multicast, where many satellites and hosted VMs may participate in the same in-orbit data distribution task. As the Starlink evaluation exhibits qualitatively similar trends to those observed in the OneWeb topology, the corresponding results are omitted in the interest of conciseness. To evaluate BIER-DC sensitivity to SDC cluster size, we select a common set of destination satellites and repartition the constellation using 2 × 3, 3 × 4, 4 × 5, and 5 × 6 clusters. We then multicast to randomly selected subsets (20\%–100\%) of the same destination set, with each subset reused across all cluster sizes. As shown in Fig.~\ref{fig:cluster_size_sensitivity}, the 2 × 3 case has the highest header overhead, as the same destinations are distributed across more SDCs, requiring more SDC-level information in the packet header.

\begin{figure*}[!t]
    \centering

    \subfloat[20\% destination density]{
        \includegraphics[width=0.28\textwidth]{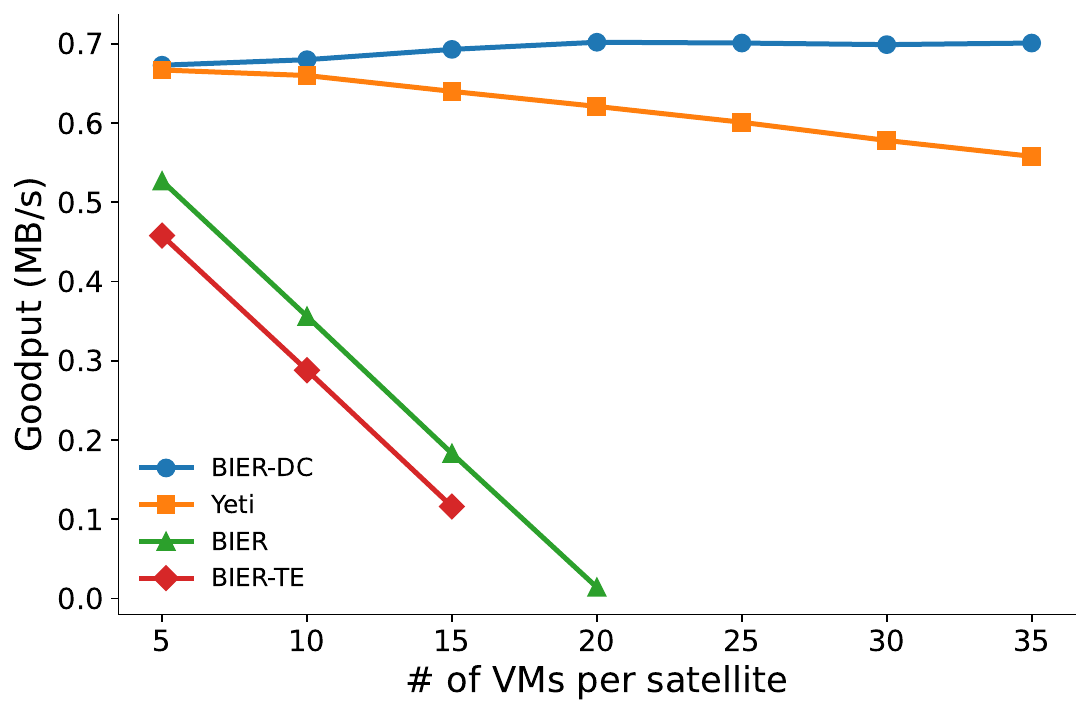}
        \label{fig:goodput-20}
    }
    \hfill
    \subfloat[40\% destination density]{
        \includegraphics[width=0.28\textwidth]{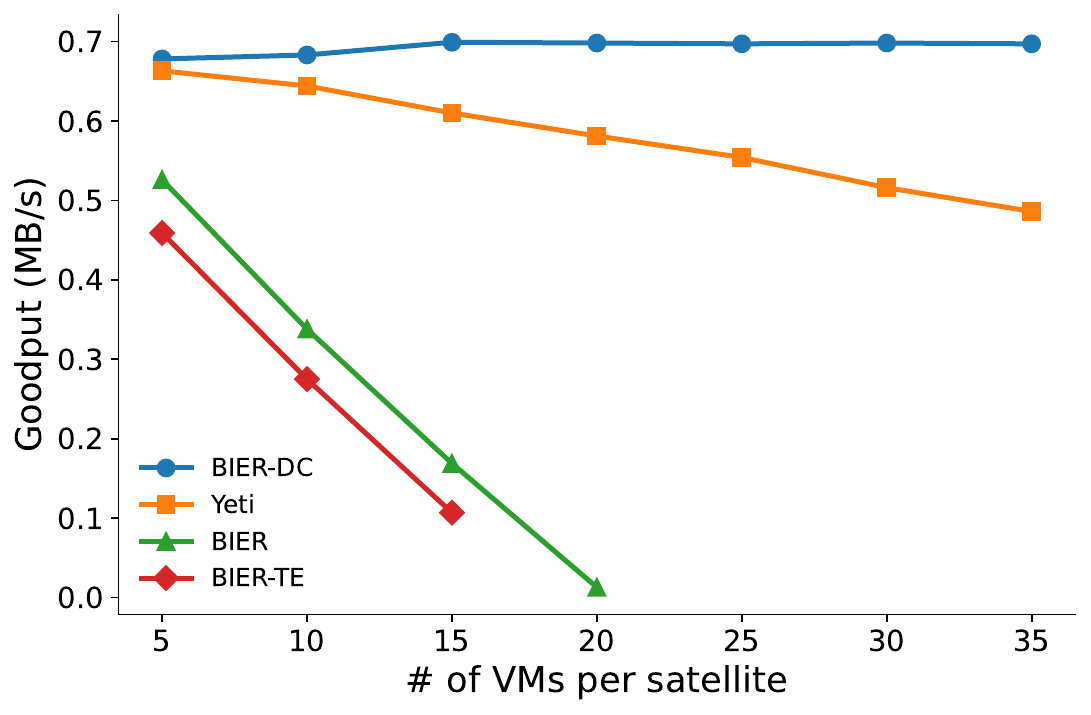}
        \label{fig:goodput-40}
    }
    \hfill
    \subfloat[60\% destination density]{
        \includegraphics[width=0.28\textwidth]{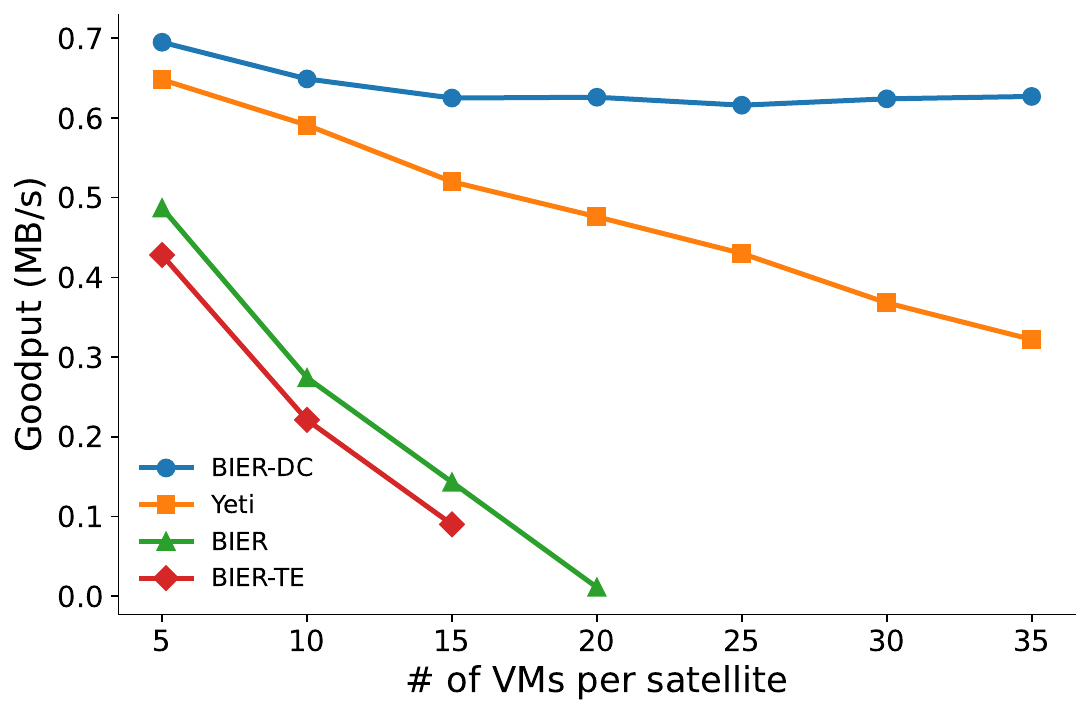}
        \label{fig:goodput-60}
    }

    \caption{Goodput versus VM count in OneWeb at destination densities of 20\%, 40\%, and 60\%. }
    \label{fig:goodput-vm-density}
\end{figure*}

\begin{figure*}[!t]
    \centering

    \subfloat[20\% destination density]{
        \includegraphics[width=0.28\textwidth]{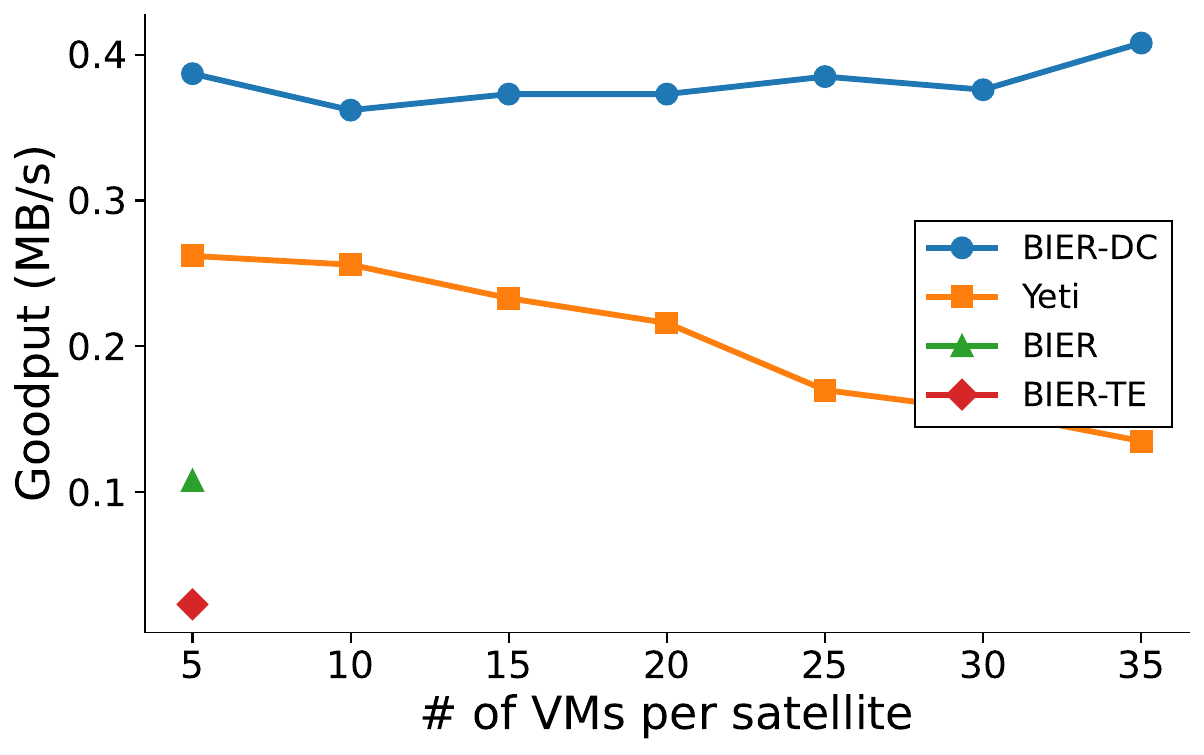}
        \label{fig:starlink-goodput-20}
    }
    \hfill
    \subfloat[40\% destination density]{
        \includegraphics[width=0.28\textwidth]{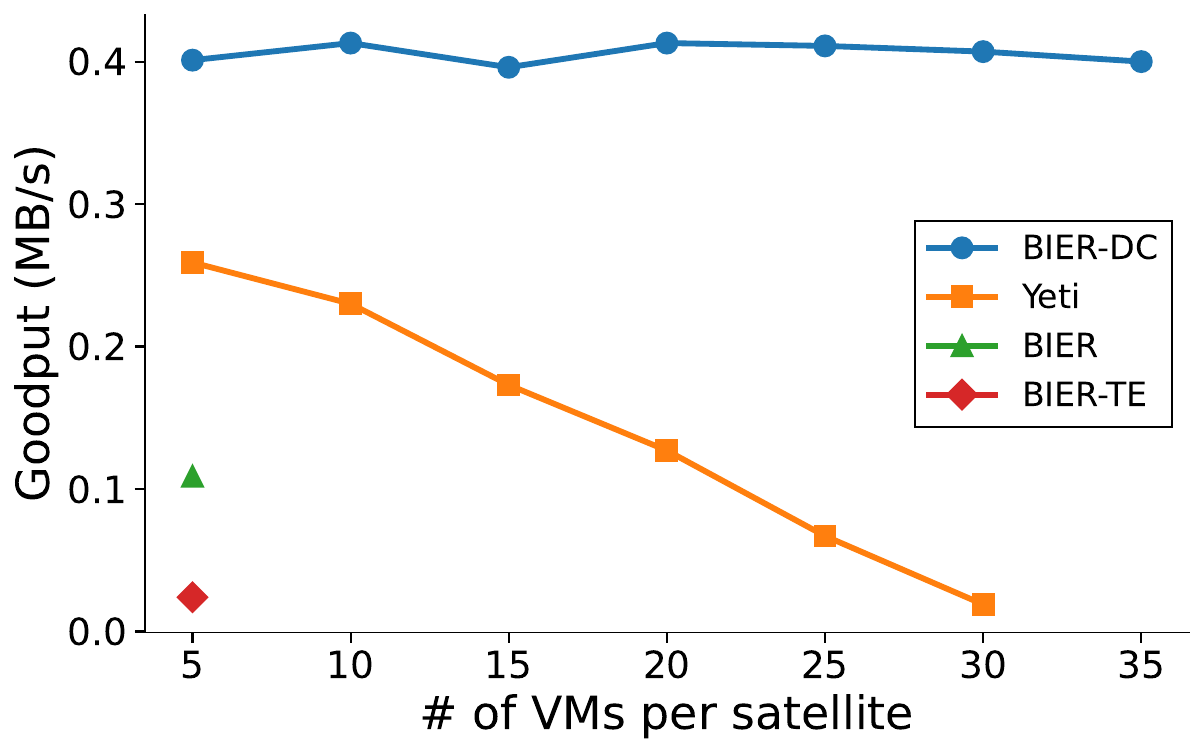}
        \label{fig:starlink-goodput-40}
    }
    \hfill
    \subfloat[60\% destination density]{
        \includegraphics[width=0.28\textwidth]{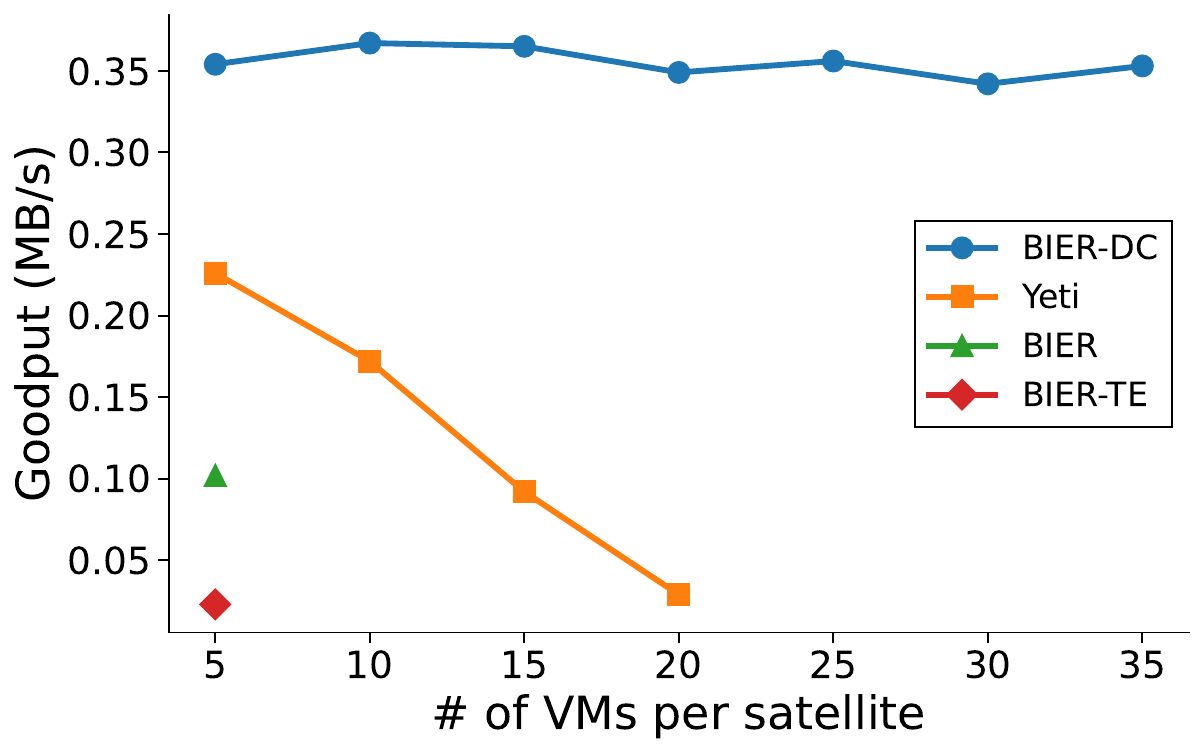}
        \label{fig:starlink-goodput-60}
    }

    \caption{Goodput versus VM count in Starlink at destination densities of 20\%, 40\%, and 60\%.}
    \label{fig:starlink-goodput-vm-density}

    \vspace{-8pt}
\end{figure*}

\noindent\textbf{Impact of Header Overhead on Goodput.} Goodput measures useful application payload delivered per unit time, excluding protocol metadata. We evaluate it by transmitting an 8~MB EO image file from one source satellite to multiple destinations at a fixed rate of 500~packets/s, with each 1500~B packet containing both protocol header and payload. For a header of $H$ bytes, only $1500-H$ bytes remain for EO data, so larger headers require more packets to deliver the same file and reduce effective goodput. For BIER and BIER-TE under larger VM configurations, the header can exceed the 1500~B packet budget, leaving no payload space and preventing the sender from forming a valid packet.

\par Fig.~\ref{fig:goodput-vm-density} and Fig.~\ref{fig:starlink-goodput-vm-density} evaluate this effect through emulation on the OneWeb and Starlink topologies under three destination-density scenarios. In all cases, one source satellite inside a source SDC multicasts the EO image file to destination satellites distributed across SDCs located 1--3 hops away from the source SDC. We vary the number of VMs per satellite on the x-axis and consider destination satellite densities of 20\%, 40\%, and 60\%, shown in Fig.~\ref{fig:goodput-20}, Fig.~\ref{fig:goodput-40}, and Fig.~\ref{fig:goodput-60}, respectively. Across all scenarios, \textsc{BIER-DC} maintains the highest and most stable goodput because its compact SDC-level header leaves most of the 1500~B packet available for useful EO payload. Yeti achieves the second-best goodput, but its performance gradually decreases as the number of VMs increases because its header consumes more packet space. BIER and BIER-TE degrade much faster: their larger headers sharply reduce the available payload per packet, and when their headers exceed the packet-size budget, the corresponding configurations become infeasible for transmission. These results show that header reduction is not only an encoding benefit, it directly improves payload efficiency and application-level goodput for SDC-based multicast.

\section{Related Work}\label{related_works_section}

\par {\bf Stateless multicast:} BIER~\cite{rfc8279} encodes the multicast receiver set as a packet-header bitstring, where each bit identifies an egress router. Routers use per-interface bitmasks to replicate and forward packets along shortest paths. BIER-TE~\cite{rfc9262} extends this model by assigning bits to adjacencies, enabling traffic-engineered multicast trees, while BIER-FRR~\cite{BIER-FRR} adds precomputed backup entries for local failure recovery. Yeti~\cite{diab2022yeti} is also stateless, using specialized labels, such as local interface IDs and interface bitmaps, in a dynamic stack that shrinks along the path. Such fine-grained encoding can still incur high header overhead in SDCs with many VMs, containers, and replicas. BIER-Star~\cite{abdollahi2026bier} improves scalability for large LEO systems by using a compact satellite-network abstraction, but it targets satellite access multicast rather than multicast among distributed computing and storage resources inside space data centers.

\par {\bf Cloud/data center multicast:} Elmo~\cite{shahbaz2019elmo} and Orca~\cite{diab2022orca} are multicast systems for terrestrial cloud/data center networks. Elmo uses a logically centralized controller to encode multicast forwarding rules in packet headers and exploits the symmetry and short paths of multi-rooted Clos fabrics. Orca targets Clos/leaf-spine data centers, using rack-level server agents and maintaining small exception state at selected spine switches. In SDCs, where each satellite acts as both a rack-like compute unit hosting multiple VMs/containers and a forwarding node connected to neighboring satellites through dynamic ISLs, directly applying Orca would require agent functionality to scale with the number of satellites and would rely on a fixed, port-based leaf/spine/core hierarchy that does not match this distributed satellite architecture. Thus, Orca is not directly suitable for space data centers.

\par BIER-DC differs from conventional BIER and hierarchical data center multicast by separating inter-SDC forwarding from intra-SDC delivery. It aggregates multiple service destinations into compact SDC–border-satellite entries and performs local header translation at destination SDCs, keeping satellite- and VM-level information out of the inter-SDC header while preserving stateless forwarding. This design avoids the fixed Clos-fabric assumptions of Elmo and Orca and is tailored to multicast among distributed compute and storage resources in dynamic LEO SDCs.

\section{Conclusion}
\label{sec:conclusion}
\par This paper presented \textsc{BIER-DC}, a compact stateless multicast design for space data centers formed by cooperative LEO satellite clusters. By encoding forwarding at the SDC level through destination-SDC and ingress-border identifiers, rather than individual satellites, VMs, or adjacencies, \textsc{BIER-DC} substantially reduces header overhead and preserves more packet capacity for application payload. Our analysis and NetSatBench-based EO image transfer emulations show that \textsc{BIER-DC} scales more efficiently and achieves higher goodput than BIER, BIER-TE, and Yeti as multicast scope and service density increase. Overall, \textsc{BIER-DC} provides a scalable multicast mechanism for in-orbit data replication, caching, and processing across distributed SDCs. Future work will extend \textsc{BIER-DC} with dynamic failure recovery, traffic-aware destination grouping, and re-evaluation of forwarding assignments over updated topology snapshots. 

\bibliographystyle{IEEEtran}
\bibliography{References.bib}

\end{document}